\documentclass[a4paper,11pt]{article}
\pdfoutput=1 

\usepackage{jheppub} 

\usepackage[T1]{fontenc} 

\usepackage{tikz}
\usetikzlibrary{chains}
\tikzset{node distance=2em, ch/.style={circle,draw,on chain,inner sep=2pt},chj/.style={ch,join},every path/.style={shorten >=4pt,shorten <=4pt},line width=1pt,baseline=-1ex}

\let\dlabel=\alabel

\newcommand{\dnode}[2][chj]{%
\node[#1,label={below:\dlabel{#2}}] {};
}

\newcommand{\dydots}{%
\node[chj,draw=none,inner sep=1pt] {\dots};
}

\title{\boldmath Quantum anomalies in $A^{(1)}_r$ Toda theories with defects }

\author[a,b]{Silvia Penati,} 
\author[c]{Davide Polvara}

\affiliation[a]{ Dipartimento di Fisica, Universit\`a degli studi di Milano--Bicocca, Piazza della Scienza 3, I-20126 Milano, Italy }
\affiliation[b]{INFN, Sezione di Milano--Bicocca, Piazza della Scienza 3, I-20126 Milano, Italy }
\affiliation[c]{Department of Mathematical Sciences, Durham University, South Road, Durham DH1 3LE, United Kingdom }

\emailAdd{silvia.penati@mib.infn.it}
\emailAdd{davide.polvara@durham.ac.uk}

\abstract{We study quantum integrability of affine Toda theories with a line of defect. In particular, we focus on the problem of constructing quantum higher-spin conserved currents in models defined by two $A_r^{(1)}$ Toda theories separated by a non-trivial type-I defect. For a suitable choice of the defect potential these theories are known to be classically integrable, that is they possess an infinite set of higher-spin conserved charges in involution. Studying the corresponding conservation laws at quantum level we discover that anomalies arise, which we compute exactly at all orders in the coupling constant.  While for the stress-energy tensor these anomalies can be cancelled by a finite renormalization of the defect potential, we find that from the first non-trivial higher-spin current this is no longer possible. This opens the question whether these theories are indeed integrable at quantum level.}

\begin{document} 
\maketitle
\flushbottom

\section{Introduction}
\label{sec:intro}

The study of quantum field theories with boundaries and defects is of primary importance in modern theoretical physics. In fact real physical systems 
have finite size, they often contain impurities, entangled systems are separated by entangling domain walls, topological phases of condensed matter can be described by boundary CFT. From a more theoretical point of view, string theory naturally contains boundaries and defects, the AdS/CFT correspondence intrinsically involve boundaries, and more generally d-dimensional theories can be studied as lying on d-dimensional domain walls of higher dimensional theories. 

The construction of new theoretical tools to unveil properties of QFT with boundary and defects is then mandatory. Recently, there has been intense activity in different directions, from the study of boundary and defect CFTs, to the renormalization group in the presence of defects, supersymmetric theories with defects, dualities, bootstrap and integrability with defects. We refer to \cite{Andrei:2018die} for a recent review and a quite exhaustive bibliography. 

In this paper we focus on the study of a particular aspect among the many, that is quantum integrability in the presence of defects. Precisely, we will be interested in investigating whether and under which conditions quantum integrable systems remain integrable after placing a codimension-one defect.  

A distinguished class of integrable models are the affine Toda theories in two dimensions. Interest in these theories is triggered not only by the fact that they are a set of exactly solvable interacting models, but also by the crucial role that they play in solving N=2 SYM theories in four dimensions \cite{Alday:2009aq,Wyllard:2009hg}.

Affine Toda theories are massive integrable field theories (IFT) owning an infinite tower of conserved charges that can be found through a Lax Pair formulation of the equations of motion \cite{a1,a2}. These charges are the natural generalisation of higher-spin light-cone components of energy and momentum. If they survive quantization the dynamics of the system is described by an S-matrix that turns out to be factorized into elastic two-particle $S$-matrices \cite{a3,a4}. 

For theories defined on the whole spatial line it has been found that every classical conservation law has its quantum version, which is simply obtained by adding suitable quantum corrections to the corresponding current \cite{a5,a6,a7,a8,a9,a10,a11}. This holds both for bosonic Toda theories \cite{a12} associated to affine Lie algebras, as well as for supersymmetric theories associated to superalgebras \cite{a13,a14,a15}. In particular, the existence of higher-spin conserved charges together with unitarity and crossing symmetry is sufficient to determine the exact $S$-matrix using a bootstrap principle \cite{a16,a17,a18,a19,a20,aa20,a21,a22,a23}. Explicit results have been found for simply laced ADE series in terms of roots and weights \cite{a24,aa24}, for the non-simply laced cases \cite{a25,a26} and for supersymetric models \cite{Delius:1990ij,Delius:1991sv}.

Over the past years the study of integrability properties has been generalized to Toda models with boundaries \cite{a27,aa27,a28}.
This has been implemented both at classical \cite{a29,a30,a31} and quantum \cite{a32,a33,a34} level. While in the classical case it is possible to select a particular class of boundary perturbations that preserve an infinite number of conserved charges, at quantum level the situation is more involved. In fact, the presence of loops gives rise to anomalous contributions that need to be cancelled by a finite renormalization of the boundary potential. This allows to preserve spin-4 current in the sinh-Gordon model and spin-3 current in $A_{r>1}^{(1)}$ models, as well as spin-4 current in nonsimply-laced Toda theories. However, strong indications arise \cite{a34} that conservation of higher-spin currents might get irreparably affected by anomalies, no matter the choice of the boundary potential. 

Lines of defects (or impurities) in two dimensional IFT have been first introduced in \cite{b15}, where it has been proved that in the presence of both reflecting and transmitting defects the only bulk to bulk $S$-matrix compatible with integrability is $S=\pm1$, i.e. the bulk theory is necessarily a free theory. The situation improves if one considers purely transmitting defects \cite{b16,b17}. In this case, for $A^{(1)}_r$ Toda systems a lagrangian formulation of the defect has been proposed \cite{b17,b18} and integrability preserving defect equations of motion has been implemented at classical level, which take the form of B\"acklund transformations frozen at the defect. Through the construction of a suitable Lax pair it has been argued that this is the only class of Toda theories which is compatible with integrability-preserving defects, as long as no extra degrees of freedom are present on the defect. Remarkably, despite the defect breaks explicitly translation symmetry in the direction orthogonal to the defect line, under suitable conditions on the form of the defect potential momentum is still conserved, and at classical level momentum conservation turns out to be sufficient for ensuring conservation of an infinite tower of higher-spin conserved charges.

Assuming that the infinite set of conserved charges survives quantization, aspects of bulk to bulk and bulk to defect scattering have been investigated both for the sine-Gordon theory \cite{b19} and for more general complex $A_r^{(1)}$ models \cite{b20}. In particular, a consistent defect transmission matrix for solitons has been proposed by solving Yang-Baxter-type equations in the region close to the defect. 

The natural question that we address in this paper is the following: "Can we reasonably assume that the classical conserved charges survive quantization, possibly under a non-trivial renormalization of the defect potential?"

Focusing on the $A_r^{(1)}$ Toda theories with a line of defect, we compute the quantum conservation laws for energy, momentum and the first higher-spin current which is conserved at classical level, that is spin-4 for sine-Gordon and spin-3 for $A_{r>1}^{(1)}$ Toda models. We apply massless perturbation theory used in \cite{a32,a33,a34} for theories with boundaries to evaluate conservation laws at all orders in the coupling constant. As already experienced in theories with boundaries, also in the present case anomalous contributions arise from quantum loops. These are {\em local} terms that affect the conservation equations in a non-trivial way. If these terms are total derivatives they simply provide quantum corrections to the defect charge. If this is not the case, the only possibility to remove them is via a finite renormalization of the defect potential, which thus acquires extra contributions proportional to higher powers in the coupling. Unfortunately we find that, apart from energy and momentum, for higher spin currents there is no defect renormalization that can cancel these anomalies and conservation laws seem to be irreparably spoiled at quantum level. 

More precisely, the main results and the structure of the paper are the following. In section \ref{sect2}, for a general 2d IFT with a line of defect we recall the classical construction of conserved charges, enlightening the constraints at the defect that need to be satisfied by the current in order to ensure charge conservation. We then describe the general approach of massless perturbation theory that we use for computing quantum corrections to the classical conservation laws and to the constraints at the defect. In section \ref{classical} we review  $A_r^{(1)}$ Toda field theories and the corresponding class of defect potentials that preserve classical integrability. For these theories, in section \ref{EM} we determine the quantum stress--energy tensor and study the energy and momentum conservation. We find that energy is automatically conserved once we take into account properly a non-trivial contribution from the defect, while momentum conservation requires a finite renormalization of the defect potential. In section \ref{higherspin} higher-spin currents are considered. In particular, we focus on the spin-4 current for the sinh-Gordon model and the spin-3 current for $A_r^{(1)}$ ($r>1$) Toda theories. In both cases we find that, in contrast with the classical situation where the conditions for momentum conservation automatically guarantee higher-spin current conservation, at quantum level the conservation laws are spoiled by anomalous contributions that cannot be cancelled through a redefinition of the defect potential. This result poses a serious question about the integrability of this class of theories, though it is not sufficient for drawing any definite conclusion. We discuss this question and possible rescuing generalizations of our results in section \ref{conclusions}. Finally, in appendix \ref{App:momentumflow} we provide a detailed example of calculation of anomalous terms, whereas appendix \ref{App:spin3} is devoted to the detailed derivation of classical conserved spin-$(\pm 2)$ charges within our approach.

\section{Conservation laws in 2d theories with defect}\label{sect2}

We consider a generic two dimensional lagrangian theory defined in the Euclidean $(x_0, x_1)$ plane, in the presence of a codimension-one defect localized at $x_1=0$. Bulk degrees of freedom propagating in the region $x_1<0$ are scalar fields $\phi^{(-)}_a$, $a=1, \dots, r$, possibly interacting with a potential $V^{(-)} \equiv V(\phi^{(-)})$, whereas in the region $x_1>0$ scalar fields $\phi^{(+)}_b$, $b=1, \dots, r$ propagate and interact via a potential $V^{(+)} \equiv V(\phi^{(+)})$. We assume that on the defect there are no further degrees of freedom, but the defect can act with non-trivial boundary conditions on the left/right fields through a localized potential $V^{(d)}(\phi^{(\pm)}, \partial_0\phi^{(\pm)})$.  
 
The theory is described by the action
\begin{equation}\begin{split}
S=& \frac{1}{\beta^2} \int_{-\infty}^{+\infty}dx_{0}\int_{-\infty}^{0}dx_{1}\biggl[\frac{1}{2}\partial_{\mu}{\phi}^{(-)}\cdot \partial_{\mu}{\phi}^{(-)}+V^{(-)} \biggr] \\ 
+ &  \frac{1}{\beta^2} \int_{-\infty}^{+\infty}dx_{0}\int_{0}^{+\infty}dx_{1}\biggl[\frac{1}{2}\partial_{\mu}{\phi}^{(+)} \cdot\partial_{\mu}{\phi}^{(+)}+V^{(+)}\biggr] \\
- & \frac{1}{\beta^2} \int_{-\infty}^{+\infty}dx_{0} \, V^{(d)}
\label{intro4}
\end{split}\end{equation}
which provides equations of motion in the bulk
\begin{equation}
\Box \phi^{(-)}_a=\frac{\partial V^{(-)}}{\partial \phi^{(-)}_a}  \hspace{4 mm},\hspace{4 mm}
\Box \phi^{(+)}_a=\frac{\partial V^{(+)}}{\partial \phi^{(+)}_a}
\label{intro4bis}
\end{equation}
supplemented by boundary conditions at the defect 
\begin{equation}
\partial_{1} \phi_a^{(-)} \Big|_{x_1=0^-}=  \frac{\partial V^{(d)}}{\partial \phi_a^{(-)}} \hspace{8 mm},\hspace{8 mm}
\partial_{1} \phi_a^{(+)} \Big|_{x_1=0^+}= - \frac{\partial V^{(d)}}{\partial \phi_a^{(+)}} 
\label{intro13bis}
\end{equation}
In the rest of this section we will assume that the two bulk theories are classically integrable, that is they allow for an infinite tower of higher spin conserved currents, and study in general how the defect interferes with the conservation laws.

\subsection{Classical currents}\label{procedure1}

We begin by reviewing the conservation laws in the two bulk theories. Working away from the defect with complex coordinates
\begin{equation}
x=\frac{1}{\sqrt{2}}(x_0+i x_1)    \hspace{8 mm}   \bar{x}=\frac{1}{\sqrt{2}}(x_0-i x_1)
\label{intro14}
\end{equation}
and corresponding derivatives
\begin{equation}
\partial \equiv \partial_x=\frac{1}{\sqrt{2}}(\partial_0-i \partial_1)    \hspace{8 mm}    \bar{\partial} \equiv \partial_{\bar{x}}=\frac{1}{\sqrt{2}}(\partial_0+i \partial_1) \hspace{8 mm} \Box=2\partial \bar{\partial}
\label{intro6}
\end{equation}
the infinite tower of higher spin conservation laws for the $(\pm)$-theories in the bulk can be written in the following way
\begin{equation}
\bar{\partial}J_{s+1}^{(\pm)} +\partial \Theta_{s-1}^{(\pm)}=0    \hspace{8 mm}    \partial \tilde{J}_{-s-1}^{(\pm)}+
\bar{\partial}\tilde{\Theta}_{-s+1}^{(\pm)}=0
\label{intro7}
\end{equation}
where the subscript indicates the spin of the current. Plus (minus) quantities are function of $\phi^+_a$ ($\phi^-_a$) fields and their derivatives, 
with $\phi^\pm_a$ satisfying the equations of motion in \eqref{intro4bis}. 

The conserved bulk charges associated to these currents at the right and left of the defect are given by 
\begin{equation}
Q_{+s}^{(\pm)}=\int_{0}^{+\infty} dx_1 \Bigl(J_{s+1}^{(\pm)}+\Theta_{s-1}^{(\pm)}\Bigr)
\quad , \quad Q_{-s}^{(\pm)}= \int_{0}^{+\infty} dx_1 \Bigl(\tilde{J}_{-s-1}^{(\pm)}+\tilde{\Theta}_{-s+1}^{(\pm)}\Bigr)  
\label{intro8.1}
\end{equation}
Equations \eqref{intro7} can be thought of as the components of a conserved vector current expressed in complex coordinates. In terms of ``time'' and ``space'' coordinates they imply two conservation laws\footnote{In order to avoid cluttering notation from now on we omit the spin label of the currents.}
\begin{equation}
\partial_0J_0^{(\pm)}+\partial_1J_1^{(\pm)}=0      \hspace{4 mm},\hspace{4 mm}    \partial_0Y_0^{(\pm)}+\partial_1Y_1^{(\pm)}=0
\label{intro8}
\end{equation}
where $J_\mu^{(\pm)}$ is the energy-like conserved current defined as
\begin{equation}
J_0^{(\pm)}=J^{(\pm)}+\Theta^{(\pm)}+\tilde{J}^{(\pm)}+\tilde{\Theta}^{(\pm)}  \hspace{8 mm}  J_1^{(\pm)}=i\Bigl(J^{(\pm)}-\Theta^{(\pm)}-\tilde{J}^{(\pm)}+\tilde{\Theta}^{(\pm)}\Bigr)
\label{intro9}
\end{equation}
and $Y_\mu^{(\pm)}$ is a momentum-like conserved current
\begin{equation}
Y_0^{(\pm)}=J^{(\pm)}+\Theta^{(\pm)}-\tilde{J}^{(\pm)}-\tilde{\Theta}^{(\pm)}  \hspace{8 mm}  Y_1^{(\pm)}=i\Bigl(J^{(\pm)}-\Theta^{(\pm)}+\tilde{J}^{(\pm)}-\tilde{\Theta}^{(\pm)}\Bigr) 
\label{intro10}
\end{equation}
The corresponding energy-like and momentum-like conserved charges in the bulk are then
\begin{equation}\begin{split}
Q_E^{(\pm)} &=\int_{\mathbb{R}^{\pm} } dx_1 J_0^{(\pm)} = Q_s^{(\pm)}  + Q_{-s}^{(\pm)}  \\
Q_P^{(\pm)} &=\int_{\mathbb{R}^{\pm}} dx_1 Y_0^{(\pm)} = Q_s^{(\pm)}  - Q_{-s}^{(\pm)} 
\label{intro11}
\end{split}\end{equation}

In order to clarify the previous definitions we consider as an example the spin-2 conserved current, that is the stress-energy tensors $T_{\mu\nu}^{(\pm)}$. In this case we can cast the conservation laws $\partial^\mu T_{\mu\nu}^{(\pm)}=0$ in the form \eqref{intro8} by setting 
\begin{equation}
 \label{eq:int_b}
 J_0^{(\pm)} = 2\, T_{00}^{(\pm)}\hspace{3mm} ,\hspace{3mm}  J_1^{(\pm)} = 2\, T_{10}^{(\pm)}  \hspace{6mm} ; \hspace{6mm}
 Y_0^{(\pm)} = -2i \, T_{01}^{(\pm)} \hspace{3mm} ,\hspace{3mm}  Y_1^{(\pm)} = -2i \, T_{11}^{(\pm)}
\end{equation}
Inverting (\ref{intro9}, \ref{intro10}) it follows that the complex components are then given by 
\begin{equation}\begin{split}
\label{eq:int_a}
& J^{(\pm)}= \frac12 (T_{00} - T_{11} - 2i T_{01}) \equiv  T_{xx}^{(\pm)} \hspace{3mm},\hspace{3mm} \tilde{J}^{(\pm)} = \frac12 (T_{00} - T_{11} + 2i T_{01}) \equiv T_{\bar{x}\bar{x}}^{(\pm)} \\ 
&\Theta^{(\pm)}=\tilde{\Theta}^{(\pm)} =  \frac12 (T_{00} + T_{11} ) \equiv  T_{x\bar{x}}^{(\pm)}=T_{\bar{x}x}^{(\pm)}
\end{split} \end{equation}
in agreement with eqs. \eqref{intro6} (we used $T_{10} = T_{01}$).

The corresponding conserved charges are the total bulk energy $\frac12 (Q_E^{(+)} + Q_E^{(-)})$ and the total momentum $\frac{i}{2}(Q_P^{(+)} + Q_P^{(-)})$, where $Q_E^{(\pm)}$ and $Q_P^{(\pm)}$ are given in \eqref{intro11}.

\vskip 10pt
So far we have worked in the bulk, away from the defect. In the presence of a defect at $x_1=0$ we expect the currents and the corresponding charges to be in general modified by a defect contribution. 
At a generic spin the total energy-like and momentum-like currents can then be written as
\begin{equation}\label{intro6bis}
\begin{split} 
J_\mu = \theta(x_1) J_\mu^{(+)} + \theta(-x_1) J_\mu^{(-)} - \delta_{\mu 0} \, \delta(x_1) \, \Sigma_E \\
Y_\mu = \theta(x_1) Y_\mu^{(+)} + \theta(-x_1) Y_\mu^{(-)} - \delta_{\mu 0} \, \delta(x_1) \, \Sigma_P 
\end{split}
\end{equation}
where $\Sigma_{E,P}$ are functions of $x_0$ localized at the defect and represent the defect contributions to the currents.
 
Using the bulk conservation laws \eqref{intro8} and taking into account that $\partial_{x_1} \theta(\pm x_1) = \pm \delta(x_1)$, it is easy to verify that the total currents in \eqref{intro6bis} satisfy
\begin{equation}
\begin{split} \label{intro6bisbis}
& \partial^\mu J_\mu = \delta(x_1) \left[ J_1^{(+)} - J_1^{(-)} - \partial_0 \Sigma_E  \right]\\
& \partial^\mu Y_\mu = \delta(x_1) \left[ Y_1^{(+)} -  Y_1^{(-)} - \partial_0 \Sigma_P  \right]
\end{split}
\end{equation}
Therefore, the conservation laws get spoiled at the defect, unless the conditions
\begin{equation}
\Bigl(J_1^{(+)}-J_1^{(-)}\Bigr)\Bigr|_{x_1=0}=\partial_0\Sigma_E\hspace{4mm},\hspace{4mm} \Bigl(Y_1^{(+)}-Y_1^{(-)}\Bigr)\Bigr|_{x_1=0}=\partial_0\Sigma_P 
\label{intro12}
\end{equation}
are satisfied. If this is the case, the corresponding total charges 
\begin{equation}
\begin{split} \label{intro12bis}
& Q_E = \int_{\mathbb{R}} dx_1 J_0 = \int_0^{+\infty} J_0^{(+)} + \int^0_{-\infty} J_0^{(-)} - \Sigma_E \\
& Q_P = \int_{\mathbb{R}} dx_1 Y_0 = \int_0^{+\infty} Y_0^{(+)} + \int^0_{-\infty} Y_0^{(-)} - \Sigma_P
\end{split}
\end{equation}
are conserved. In fact, following the usual Noether procedure, we apply the time derivative to these expressions, use the bulk conservations laws to trade $\partial_0 J_0$ with  $\partial_1 J_1$ inside the integrals and integrate by parts the spatial derivative. In the presence of finite integration boundaries, we obtain net current flows at the defect 
\begin{equation}
\partial_0Q_E=\Bigl(J_1^{(+)}-J_1^{(-)}\Bigr)\Bigr|_{x_1=0} - \partial_0 \Sigma_E \hspace{4mm},\hspace{4mm}
\partial_0Q_P=\Bigl(Y_1^{(+)}-Y_1^{(-)}\Bigr)\Bigr|_{x_1=0} - \partial_0 \Sigma_P
\label{intro11.1}
\end{equation} 
and the right hand sides of these equations vanish identically if constraints \eqref{intro12} are satisfied. 

Equivalently, we can re-express the conservation laws in terms of spin-$(\pm s)$ charges. Using relations \eqref{intro11}  and \eqref{intro12bis} the total conserved charges are given by
\begin{equation} 
\begin{split}
& Q_{+s} = Q_{+s}^{(+)} +  Q_{+s}^{(-)}  - \frac12 (\Sigma_E + \Sigma_P) \\
& Q_{-s} = Q_{-s}^{(+)} +  Q_{-s}^{(-)}  - \frac12 (\Sigma_E - \Sigma_P)
\end{split}
\end{equation} 
and the corresponding conservation laws read
\begin{equation}
\label{intro19}
\begin{split}
\partial_0Q_{+s}&=i \Bigl( J^{(+)} - J^{(-)} + \Theta^{(-)} - \Theta^{(+)} \Bigr)\Bigr|_{x_1=0} - \frac12 (\partial_0 \Sigma_E + \partial_0 \Sigma_P) =0 \\  
\partial_0Q_{-s}&=i \Bigl( \tilde{J}^{(-)} - \tilde{J}^{(+)} + \tilde{\Theta}^{(+)} - \tilde{\Theta}^{(-)} \Bigr)\Bigr|_{x_1=0} - \frac12 (\partial_0 \Sigma_E - \partial_0 \Sigma_P)  =0
\end{split}
\end{equation} 
where the r.h.s. vanish thanks to constraints \eqref{intro12}. 

When applied to the energy-momentum tensor, identities \eqref{intro6bisbis} specialize to
\begin{equation} 
\partial^\mu T_{\mu 1} = \delta(x_1) \left[ T_{11}^{+} - T_{11}^{-} - \partial_0 \Sigma_P \right] \equiv \delta(x_1) D(x_0)
\end{equation} 
where $D(x_0)$ is nothing but the displacement operator, which signals the breaking of momentum conservation along the direction transversal to the defect.

In conclusion, in the presence of a line of defects classical integrability survives if the bulk conservation laws \eqref{intro8} are supplemented by boundary conditions \eqref{intro12}. This generalizes what happens in two-dimensional theories with a boundary \cite{a29,a30,a31,a32,a33,a34}. In fact, the present case reduces to that one if we set one type of fields to zero, for instance $\phi^{(-)}_a=0$ for any $a$.

\subsection{Quantum currents} \label{procedure}

We are interested in determining the general class of defects that preserve integrability at quantum level. This amounts to study quantum corrections to conservation laws (\ref{intro7}) (or equivalently \eqref{intro8}) and constraints (\ref{intro12}).

We use the technique of massless perturbation theory that allows to find exact results to all orders in the $\beta^2$ coupling \cite{a12,a13,a14,a15,a32,a33,a34}.
This consists in writing the action in (\ref{intro4}) in the following form
\begin{equation}
S=S_0+S^{(-)}_V+S^{(+)}_V-S_d
\label{sec2_1}
\end{equation}
where $S_0$ describes two free theories defined on the left and on the right regions of the spatial axis
\begin{equation}
S_0=\frac{1}{2\beta^2}\int_{-\infty}^{+\infty}dx_{0}\int_{-\infty}^{0}dx_{1} \, \partial_{\mu}{\phi}^{(-)} \cdot  \partial_{\mu}{\phi}^{(-)} + \frac{1}{2\beta^2}\int_{-\infty}^{+\infty}dx_{0}\int_{0}^{+\infty}dx_{1} \, \partial_{\mu}{\phi}^{(+)} \cdot \partial_{\mu}{\phi}^{(+)}
\label{sec2_2}
\end{equation}
while $S^{(\pm)}_V = \frac{1}{\beta^2} \int_\mathbb{R} dx_0 \int_{\mathbb{R}^{\pm}} dx_1: \hspace{-0.1cm}V^{(\pm)} \hspace{-0.1cm} :$  and $S_d = \frac{1}{\beta^2} \int_\mathbb{R} dx_0 : \hspace{-0.1cm} V^{(d)} \hspace{-0.1cm} :$, which are respectively the bulk and defect interactions, are treated as perturbations \footnote{We normal order the interaction terms in order to avoid UV divergences.}.
The non-vanishing propagators from the free action in (\ref{sec2_2}), satisfying the free boundary conditions $\partial_1 G^{(\pm)}_{ab} \Big|_{x_1=0^{\pm}} \hspace{-0.5cm}=0$, are given by 
\begin{equation}
G_{ab}^{(\pm)}(x,y) = - \delta_{ab}  \, \frac{\beta^2}{4\pi}\Bigl[\log2|x-y|^2+\log2|x-\bar{y}|^2\Bigr]
\label{sec2_3}
\end{equation}

The quantum counterparts of conservation laws \eqref{intro7} and flow conservation through the defect \eqref{intro12} can be determined by evaluating the vacuum expectation values 
\begin{equation} \label{sec2_5tris}
\bar{\partial}\bigl \langle J^{(\pm)}(x) \bigr \rangle \equiv \bar{\partial}\bigl \langle J^{(\pm)}(x) e^{-S_V^{(\pm)}}\bigr \rangle_0 \qquad , \qquad \partial\bigl \langle \tilde{J}^{(\pm)}(x) \bigr \rangle \equiv \bar{\partial}\bigl \langle \tilde{J}^{(\pm)}(x) e^{-S_V^{(\pm)}}\bigr \rangle_0 
\end{equation}
and
\begin{equation}
\Bigl \langle J_1^{(+)}-J_1^{(-)}\Bigr \rangle\Bigr|_{x_1=0} \equiv \Bigl \langle \bigl(J_1^{(+)}-J_1^{(-)}\bigr) e^{-S_V^{(+)} - S_V^{(-)} + S_d}\Bigr \rangle_0 \Bigr|_{x_1=0}
\label{sec2_5bis}
\end{equation}
To this end, we split all the fields in the action and the currents as $\phi^{(\pm)} \to \phi^{(\pm)} + \phi_q^{(\pm)}$, where $\phi^{(\pm)}$ are the classical fields satisfying free equations of motion, $\partial \bar{\partial} \phi^{(\pm)} = 0$,  and trivial boundary conditions $\partial_1 \phi^{(\pm)}|_{x_1= 0^{\pm}}=0$, whereas $\phi_q^{(\pm)}$ are the quantum fluctuations around them. We then use Wick theorem with free propagators \eqref{sec2_3} to contract the quantum fields of the currents with fields in the expansion of the interaction actions. 

In general, quantum corrections to the conservation laws in the bulk and at the defect will take the form (we restrict the discussion to energy-like currents, as for momentum-like ones the procedure is identical)
\begin{equation}\begin{split}
\label{sec2_5}
& \bar{\partial}\bigl \langle J^{(\pm)}(x) \bigr \rangle =-\partial \bigl \langle \Theta^{(\pm)}(x) \bigr \rangle + \text{non-$\partial$ terms} \\
& \partial\bigl \langle \tilde{J}^{(\pm)}(x) \bigr \rangle =-\bar{\partial} \bigl \langle \tilde{\Theta}^{(\pm)}(x) \bigr \rangle + \text{non-$\bar{\partial}$ terms} 
\end{split}
\end{equation}
\begin{equation}
\Bigl \langle J_1^{(+)}-J_1^{(-)}\Bigr \rangle\Bigr|_{x_1=0}  = \; \partial_0 \Sigma_E  + \text{non-$\partial_0$ terms}
\label{sec2_9}
\end{equation}
Here ''non-$\partial, \bar{\partial}$ terms'' and ''non-$\partial_0$ terms'' indicate possible anomalous contributions. These are {\em local} terms that cannot be absorbed by a redefinition of $\Theta^{(\pm)}, \tilde{\Theta}^{(\pm)}$ and $\Sigma_E$ and, if present, spoil the conservation law at quantum level. We will then look for conditions on the structure of the current and the defect potential which ensure the vanishing of these quantum anomalies. 

Precisely, the procedure goes as follows. We begin by writing the most general expression for the complex components of spin-$(s+1)$ currents in $x_1>0$ ($J^{(+)}, \tilde{J}^{(+)}$) and $x_1<0$ ($J^{(-)}, \tilde{J}^{(-)}$) 
\begin{equation}
J^{(\pm)}=\sum_{n=1}^{s+1}\sum_{ab} C_{ab} \, \partial^{a_1}\phi_{b_1}^{(\pm)}\ldots\partial^{a_n}\phi_{b_n}^{(\pm)} \quad , \quad \tilde{J}^{(\pm)}=\sum_{n=1}^{s+1}\sum_{ab} \bar{C}_{ab} \, \bar{\partial}^{a_1}\phi_{b_1}^{(\pm)}\ldots \bar{\partial}^{a_n}\phi_{b_n}^{(\pm)} 
\label{sec2_4}
\end{equation}
where $a\equiv(a_1,..,a_n)$, $b\equiv(b_1,..,b_n)$, $1 \leq a_j \leq s+1$ and $\sum_{j=1}^n a_j=s+1$. In general, the $C_{ab}$ coefficients can be chosen to be real, $\bar{C}_{ab} = C_{ab}$, and are given by a power expansion in the coupling constant
\begin{equation}
C_{ab}=C^{(0)}_{ab}+\beta^2 C^{(1)}_{ab}+\ldots
\label{sec2_6}
\end{equation}
Here $C^{(0)}_{ab}$ is the classical coefficient determined by imposing the classical conservation equations $\bar{\partial}J^{(\pm)}=-\partial \Theta^{(\pm)}$. 

The quantum corrections in \eqref{sec2_6} are obtained by evaluating \eqref{sec2_5tris} and imposing the anomalous terms to cancel.  Expanding the interaction potential $e^{-S^{(\pm)}_V}$ at first order and contract quantum fields of the currents with quantum fields appearing in  $S^{(\pm)}_V$, local contributions arise from terms of the form 
\begin{equation}
 \bar{\partial}\bigl \langle J^{(\pm)}(x)\,  S_V^{(\pm)} \bigr \rangle_0 \longrightarrow  
\, \bar{\partial}_x \hspace{-0.1cm} \int_{-\infty}^{+\infty} d\omega_0 \int_{\mathbb{R}^{\pm}} d\omega_1 \, \mathcal{M}_k^{(\pm)}(x,\bar{x})\Bigl[ \frac{1}{(x-\omega)^k}+\frac{1}{(x-\bar{\omega})^k} \Bigr] \mathcal{N}_k^{(\pm)}(\omega,\bar{\omega})
\label{sec2_7}
\end{equation}
where $\mathcal{M}_k^{(\pm)}$ and $\mathcal{N}_k^{(\pm)}$ correspond to the classical fields of the current and of the interaction potential, respectively, which do not undergo contractions. Here $k$ is an integer that cannot exceed the spin of the current. Local expressions arise from \eqref{sec2_7} by using the general identity 
\begin{equation}
\bar{\partial}_x \frac{1}{(x-\omega)^k}=\frac{2 \pi}{(k-1)!} \partial^{k-1}_{\omega} \delta^{(2)}(x-\omega).
\label{sec2_8}
\end{equation}
Since $\bar{\omega}$ does not belong to the half plane over which we are integrating, the term $\frac{1}{(x-\bar{\omega})^k}$ never provides local contributions. 

The first order expansion in $S_V^{(\pm)}$ does exhaust all possibilities to obtain local terms. In fact, although for a spin-$(s+1)$ current non--vanishing contributions come from contractions with all powers $(S^{(\pm)}_V)^p$, with $1 \leq p \leq s+1$, contractions with $p>1$ powers lead to $p>1$ $\omega$-integrations. On the other hand, we have only one $\bar{\partial}$-derivative in the game, which implies only one delta function produced by identity \eqref{sec2_8} and therefore only one $\omega$-integration duable. It follows that all the contributions with $p>1$ are necessarily non-local and we can drop them. 
  
If local contributions obtained in this way can be written as $\partial$-derivative of some local expression they provide quantum corrections to the classical 
traces $\Theta^{(\pm)}$, according to \eqref{sec2_5}. If not, then they have to be removed by a finite renormalization of the classical current. This procedure determines the quantum structure of $J^{(\pm)}$ and $\Theta^{(\pm)}$ exactly, to all orders in the $\beta^2$ coupling. The same procedure can be used to determine the quantum antiholomorphic conuterparts, $\tilde{J}^{(\pm)}$ and $\tilde{\Theta}^{(\pm)}$.

Armed with the exact expressions for the quantum bulk currents, we proceed looking for the conditions that preserve the flow through the defect. This amounts to study the vacuum expectation value \eqref{sec2_5bis}, where we expand $e^{-S_V^{(+)} - S_V^{(-)} + S_d}$ in powers of the potentials. 

There, we evaluate $\langle J_1^{(+)} \rangle\Bigr|_{x_1=0}$ and $ \langle J_1^{(-)} \rangle\Bigr|_{x_1=0}$separately, writing
\begin{equation}
\Bigl \langle J_1^{(\pm)}\Bigr \rangle\Bigr|_{x_1=0} = i \Bigl\langle \left[ (J^{(\pm)} - \Theta^{(\pm)}) - (\tilde{J}^{(\pm)} - \tilde{\Theta}^{(\pm)}) \right] e^{-S_V^{(+)} - S_V^{(-)} + S_d}\Bigr\rangle_0\Bigr|_{x_1=0}
\end{equation} 
In both cases, contractions with the bulk potentials generate classical contributions that could be equivalently obtained by using interacting equations of motion \eqref{intro4bis} for the classical fields. Instead, from contractions with the first order term $S_d$ contributions may arise which have the following general structure
\begin{equation}
\lim_{x_1 \to 0^\pm} \int_{-\infty}^{+\infty} \hspace{-0.5cm} d\omega_0 \biggl[\mathcal{P}^{(\pm)}(x,\bar{x}) \biggl(\frac{1}{(x-\omega)^k}+\frac{1}{(x-\bar{\omega})^k} \biggr) -\tilde{\mathcal{P}}^{(\pm)}(x,\bar{x})\biggl(\frac{1}{(\bar{x}-\bar{\omega})^k}+\frac{1}{(\bar{x}-\omega)^k}\biggr) \biggl] \mathcal{Q}(\omega_0)
\label{sec2_10}
\end{equation}
where $\mathcal{P}^{(\pm)}$ ($\tilde{\mathcal{P}}^{(\pm)}$) and $\mathcal{Q}$ correspond respectively to the part of the $(J^{(\pm)} -\Theta^{(\pm)})$ ($(\tilde{J}^{(\pm)} - \tilde{\Theta}^{(\pm)})$) currents and the part of the defect interaction that have not been contracted. In particular, $\mathcal{P}^{(\pm)}$ are functions of $\partial^p \phi^{(\pm)}$, whereas $\tilde{\mathcal{P}}^{(\pm)}$ are the same functions but with $\partial \to \bar{\partial}$. 

Since at the defect $\omega = \bar{\omega} = \frac{1}{\sqrt{2}}\omega_0$, the above expression is proportional to
\begin{equation}
\lim_{x_1 \to 0^\pm} \int_{-\infty}^{+\infty} \hspace{-0.3cm} d\omega_0 \biggl[\mathcal{P}^{(\pm)}(x,\bar{x}) \frac{1}{(x_0-\omega_0+ix_1)^k}  -\tilde{\mathcal{P}}^{(\pm)}(x,\bar{x})\frac{1}{(x_0-\omega_0-ix_1)^k} \biggl] \mathcal{Q}(\omega_0)
\label{sec2_11}
\end{equation}
Therefore, local defect contributions arise from terms in $\mathcal{P}|_{x_1=0}$ and $\tilde{\mathcal{P}}|_{x_1=0}$ which are identical. In fact, pulling out this common part and using the following identity
\begin{equation}
\lim_{x_1 \to 0^\pm} \biggl(\frac{1}{(x_0-\omega_0+ix_1)^k} - \frac{1}{(x_0-\omega_0-ix_1)^k}\biggr)=\mp \frac{2 \pi i }{(k-1)!} \partial^{k-1}_{\omega_0}\delta(x_0-\omega_0)
\label{sec2_12}
\end{equation}
we can perform the $\omega_0$ integration in \eqref{sec2_11}, so landing on a local expression. 

It is important to stress that in order to be able to use identity \eqref{sec2_12} it is crucial that in \eqref{sec2_11} the relative sign between the two integrands is minus. For the energy-like currents under discussion, at first order in $S_d$ this is guaranteed, as we evaluate $\langle (J^{(\pm)} -\Theta^{(\pm)}) \rangle$ minus $\langle(\tilde{J}^{(\pm)} - \tilde{\Theta}^{(\pm)})\rangle$.
For the momentum-like currents, instead, we have to evaluate $\langle (J^{(\pm)} -\Theta^{(\pm)}) \rangle$ plus $\langle(\tilde{J}^{(\pm)} - \tilde{\Theta}^{(\pm)})\rangle$ (see eq. \eqref{intro10}), so that applying the same procedure we end up with an expression of the form (\ref{sec2_11}) but with a relative plus sign between the two integrands. Therefore identity (\ref{sec2_12}) cannot be used and we conclude that at the first order in $S_d$ we do not obtain local contributions. In this case the first local contributions arise at second order, from contractions with $S_d^2$.

In general, it can be proved that local contributions to the energy-like part of the current arise from odd powers in $S_d$,
while contributions to the momentum-like current are given by even powers in $S_d$. We will give evidence of this statement in the explicit examples that we are going to investigate in the rest of the paper. An explicit calculation where contributions come from contractions with $S^2_d$ is given in appendix \ref{App:momentumflow}. 

Every time the number of delta functions produced by repeated application of identity \eqref{sec2_12} equals the number of integrations we obtain local terms, hence potential anomalies at the defect. If these terms can be cast in the form $\partial_0$(something) by a finite renormalization of the defect potential the conservation laws in the presence of a defect can be restored at quantum level. Otherwise, the defect spoils the conservation law.  

We conclude this section with an important observation concerning the role of total derivatives in the currents \cite{a32,a33,a34}.
It is well-known that at classical level bulk currents are always defined up to total derivatives. In fact, the conservation laws in \eqref{intro7} are trivially invariant under transformations
\begin{equation}\begin{split}
\label{sec2_14}
&J^{(\pm)} \rightarrow J^{(\pm)}+\partial{U^{(\pm)}} \qquad \qquad \Theta^{(\pm)} \rightarrow \Theta^{(\pm)}-\bar{\partial}{U^{(\pm)}}\\
&\tilde{J}^{(\pm)} \rightarrow \tilde{J}^{(\pm)}+\bar{\partial}{\tilde{U}^{(\pm)}} \qquad \qquad \tilde{\Theta}^{(\pm)} \rightarrow \tilde{\Theta}^{(\pm)}-\partial{\tilde{U}^{(\pm)}}\\
\end{split}\end{equation}
These correspond to changing the spatial components of the energy-like and momentum-like quantities by a $\partial_0$-derivative
 \begin{equation}\begin{split}
 \label{sec2_15}
& J_1^{(\pm)} \rightarrow J_1^{(\pm)} + i\sqrt{2} \partial_0(U^{(\pm)}-\tilde{U}^{(\pm)})\\
&Y_1^{(\pm)} \rightarrow Y_1^{(\pm)} + i\sqrt{2} \partial_0(U^{(\pm)}+\tilde{U}^{(\pm)})  
\end{split} \end{equation}
so that the constraints at the defect \eqref{intro12} are only altered by a shift of the $\Sigma_{E,P}$ quantities. 

At quantum level the situation is very different. In fact, if we consider loop corrections to $\bar{\partial} \langle \partial U^{(i)}\rangle$ new terms can be generated in the quantum trace that cannot be expressed as $\bar{\partial}$(something). As a consequence, they modify the currents in \eqref{sec2_15} 
by quantities that are not automatically in the form of $\partial_0$-derivatives, so providing more freedom in the search for solutions to the defect constraints. For this reason we will consider the possibility to modify the currents with total $\partial$-derivative terms. \\

\section{$A_r^{(1)}$ Toda theories with defect: A review of classical integrability}\label{classical}

The $A_r^{(1)}$ Toda theory is a two-dimensional theory of $r$ scalar fields, $\phi=(\phi_1,\ldots,\phi_r)$, interacting through the exponential potential 
\begin{equation}
V=\sum_{j=0}^r e^{\alpha_j \cdot \phi}
\label{intro1}
\end{equation}
where $\{\alpha_j\}_{j=1}^r$ are the simple roots of the Lie algebra $su(r+1)_{\mathbb{C}}$ and $\alpha_0=-\sum_{j=1}^r \alpha_j$. Assuming that 
$\alpha_i^2=2$, $i=1,..,r$ the scalar product between two roots is given by
\begin{equation}
\alpha_i \cdot \alpha_j = 2 \delta_{i j}- \delta_{i, j+1}-\delta_{i, j-1}
\label{intro2}
\end{equation}
The corresponding affine Dynkin diagram associated to this algebra is represented by the graph below
\begin{align*}
A_r^{(1)} (r \ge 1)&&&
\begin{tikzpicture}[start chain,node distance=1ex and 2em]
\dnode{1}
\dnode{2}
\dydots
\dnode{r-1}
\dnode{r}
\begin{scope}[start chain=br going above]
\chainin(chain-3);
\node[ch,join=with chain-1,join=with chain-5,label={[inner sep=1pt]10:\(\alpha_0\)}] {};
\end{scope}
\end{tikzpicture}
\end{align*}
When defined on the whole plane, these theories are classically and quantum integrable \cite{a1,a2,a12,a13,a14,a15,a16,a17,a18,a19,a20,aa20,a21,a22,a23,a24,aa24,a25,a26,Delius:1990ij,Delius:1991sv}. 

We consider a defect field theory described by the action in \eqref{intro4} where now the potential $V^{(\pm)}$ is the Toda potential \eqref{intro1} written in terms of the $\phi^{(-)}$ multiplet for $x_1<0$ and the $\phi^{(+)}$ one for $x_1>0$ \footnote{Without loosing generality, we can take the same root basis at the left and the right of the defect, as a suitable rotation of the fields can be always performed to realize this configuration.}.
Moreover, following previous literature \cite{b17,b18}  we choose to sew the two sets of fields at the defect by a potential of the form
 \begin{equation}
 V^{(d)}= \frac{i}{2} \phi_a^{(-)}E_{ab}\partial_{0}\phi_b^{(-)}+i\phi_a^{(-)}D_{ab}\partial_{0}\phi_b^{(+)}+\frac{i}{2}\phi_a^{(+)}F_{ab}\partial_{0}\phi_b^{(+)}+B(\phi^{(-)},\phi^{(+)})
 \label{intro3}
 \end{equation}
where $D, E$ and $F$ are constant matrices and, neglecting total time derivatives in the lagrangian, $E$ and $F$ can be taken antisymmetric. The scalar potential $B$ is a function of the fields but not of their derivatives. 
 
The bulk equations of motion are \eqref{intro4bis}, whereas the conditions at the defect are Backlund-type trasformations explicitly given by
\begin{equation}\begin{split}
&\partial_{1} \phi_a^{(-)}\Big|_{x_1 = 0^-}=iE_{ab}\partial_{0}\phi_b^{(-)} + iD_{ab} \partial_0 \phi^{(+)}_b + \frac{\partial B}{\partial \phi_a^{(-)}}\\
&\partial_{1} \phi_a^{(+)}\Big|_{x_1 = 0^+}=-iF_{ab}\partial_{0}\phi_b^{(+)} + iD_{ba} \partial_0 \phi^{(-)}_b - \frac{\partial B}{\partial \phi_a^{(+)}} 
\label{intro13}
\end{split}\end{equation}

It has been proved \cite{b18} that for a suitable choice of the $E$, $D$ and $F$ matrices and the $B$ potential, the corresponding defect Toda theory is classically integrable. 

\vskip 10pt
Precisely, for $r=1$ the action describes the sinh-Gordon model. Following the convention in (\ref{intro2}) the roots are given by $\alpha_0=-\alpha_1=-\sqrt{2}$ and the bulk potentials on the left and the right of the defect read
\begin{equation}
V^{(\pm)}=e^{\sqrt{2}\phi^{(\pm)}}+e^{-\sqrt{2}\phi^{(\pm)}} 
\label{sec31_1}
\end{equation}
In the presence of the defect the model possesses an infinite set of spin-even\footnote{Invariance of the theory under $\phi^{(\pm)} \to - \phi^{(\pm)}$ forces the spin of the conserved currents to be even.} conserved currents if $D=1$, $E=F=0$ and 
\begin{equation}
B=\frac{1}{\sigma} \Bigl( e^{\frac{1}{\sqrt{2}} ( \phi^{(+)}+\phi^{(-)})} +e^{\frac{1}{\sqrt{2}} ( -\phi^{(+)}-\phi^{(-)})}\Bigr)+ \sigma \Bigl( e^{\frac{1}{\sqrt{2}} (\phi^{(+)}-\phi^{(-)})}+e^{\frac{1}{\sqrt{2}} (-\phi^{(+)}+\phi^{(-)})} \Bigr)
\label{sec31_4}
\end{equation}
with $\sigma$ being a free constant parameter. In particular, beyond the stress-energy tensor the first non-trivial classically conserved current is a spin-4 current with complex components 
\begin{equation} \begin{split}\label{intro13bis}
& J^{(\pm)} =  ( \partial \phi^{(\pm)} )^4 + 2 ( \partial^2 \phi^{(\pm)} )^2 \qquad , \qquad   \Theta^{(\pm)} = - \frac{\partial^2 V^{(\pm)}}{\partial {\phi^{(\pm)}}^2} (\partial \phi^{(\pm)})^2  \\
& \tilde{J}^{(\pm)} =  ( \bar{\partial} \phi^{(\pm)} )^4 + 2 ( \bar{\partial}^2 \phi^{(\pm)} )^2  \qquad , \qquad   \tilde{\Theta}^{(\pm)} = - \frac{\partial^2 V^{(\pm)}}{\partial {\phi^{(\pm)}}^2} (\bar{\partial} \phi^{(\pm)})^2 
\end{split}\end{equation}
satisfying eq. \eqref{intro7}. 
The corresponding energy-like and momentum-like currents of the form \eqref{intro6bis} are both conserved with
\begin{equation}\begin{split}
&\Sigma_E = \frac{1}{\sigma^3} \biggl[ -6 \cosh \Bigl[\frac{1}{\sqrt{2}} (\phi^{(-)}+\phi^{(+)})\Bigr] + \frac{2}{3} \cosh \Bigl[\frac{3}{\sqrt{2}} (\phi^{(-)}+\phi^{(+)}) \Bigr] \biggr]\\
&+ \frac{1}{\sigma} \biggl[ 12 \cosh \Bigl[\frac{1}{\sqrt{2}} (\phi^{(-)}-\phi^{(+)}) \Bigr] - 6 \cosh \Bigl[\frac{1}{\sqrt{2}} (3\phi^{(-)}+\phi^{(+)}) \Bigr]-6 \cosh \Bigl[\frac{1}{\sqrt{2}} (\phi^{(-)}+3\phi^{(+)}) \Bigr] \biggr]\\
&+\sigma \biggl[ 12 \cosh \Bigl[\frac{1}{\sqrt{2}} (\phi^{(-)}+\phi^{(+)}) \Bigr] - 6 \cosh \Bigl[\frac{1}{\sqrt{2}} (\phi^{(-)}-3\phi^{(+)}) \Bigr]-6 \cosh \Bigl[\frac{1}{\sqrt{2}} (\phi^{(+)}-3\phi^{(-)}) \Bigr] \biggr]\\
&+\sigma^3 \biggl[ -6 \cosh \Bigl[\frac{1}{\sqrt{2}} (\phi^{(-)}-\phi^{(+)})\Bigr] + \frac{2}{3} \cosh \Bigl[\frac{3}{\sqrt{2}} (\phi^{(-)}-\phi^{(+)}) \Bigr] \biggr]
\end{split}\end{equation} 
and
\begin{equation}\begin{split}
&\Sigma_P =\frac{1}{\sigma^3} \biggl[ - 6 \cosh \Bigl[\frac{1}{\sqrt{2}} (\phi^{(-)}+\phi^{(+)})\Bigr] + \frac{2}{3} \cosh \Bigl[\frac{3}{\sqrt{2}} (\phi^{(-)}+\phi^{(+)}) \Bigr] \biggr]\\
&+\frac{1}{\sigma} \biggl[ 4 \cosh \Bigl[\frac{1}{\sqrt{2}} (\phi^{(-)}-\phi^{(+)}) \Bigr] - 2 \cosh \Bigl[\frac{1}{\sqrt{2}} (3\phi^{(-)}+\phi^{(+)}) \Bigr]- 2 \cosh \Bigl[\frac{1}{\sqrt{2}} (\phi^{(-)}+3\phi^{(+)}) \Bigr] \biggr]\\
&+\sigma \biggl[ - 4 \cosh \Bigl[\frac{1}{\sqrt{2}} (\phi^{(-)}+\phi^{(+)}) \Bigr] + 2  \cosh \Bigl[\frac{1}{\sqrt{2}} (\phi^{(-)}-3\phi^{(+)}) \Bigr]+ 2 \cosh \Bigl[\frac{1}{\sqrt{2}} (\phi^{(+)}-3\phi^{(-)}) \Bigr] \biggr]\\
&+\sigma^3 \biggl[ 6 \cosh \Bigl[\frac{1}{\sqrt{2}} (\phi^{(-)}-\phi^{(+)})\Bigr] - \frac{2}{3}  \cosh \Bigl[\frac{3}{\sqrt{2}} (\phi^{(-)}-\phi^{(+)}) \Bigr] \biggr]
\end{split}\end{equation} 
It is important to recall that the chosen defect potential ensures energy and momentum conservation as defined in \eqref{intro12bis}, though the defect breaks translational invariance along $x_1$ \cite{b17,b18}. This is equivalent to the conservation of both $Q_1^{(\pm)}$ and $Q_{-1}^{(\pm)}$ (see eq. \eqref{intro11}) or any linear combination of the two.

\vskip 10pt
For $r>1$ the model possesses an infinite set of conserved currents of any spin $s +1\geq 2$ if the $E,F,D$ matrices are for instance given by (for details, see \cite{b21})
\begin{equation} \label{sec2_14bis}
 E =  F = 1 - D = \sum_{i=1}^r \left( \lambda_{i+1} \lambda_{i}^T - \lambda_{i} \lambda_{i+1}^T \right)   
\end{equation}
where $\lambda_i$ are the fundamental weights of the $A_r^{(1)}$ algebra ($\alpha_i \cdot \lambda_j = \delta_{ij}$, $\lambda_0 = 0$), and the defect potential is chosen to be
\begin{equation}
B=\sum_{i=0}^r\Bigl[ \frac{1}{\sigma} \, e^{\frac12 \alpha_i \cdot R}+\sigma \, e^{\frac{1}{2}\alpha_i \cdot  S} \Bigr]
\label{sec21_14}
\end{equation}
with $\sigma$ being still a free parameter and, for later convenience, we have introduced the combinations of fields 
\begin{equation}
R \equiv  D \phi^{(+)}  + D^T \phi^{(-)}   \qquad , \qquad  S \equiv D(\phi^{(-)}-\phi^{(+)})    
\label{sec321_5}
\end{equation} 

We recall that the $B$ potential satisfies the condition
\begin{equation}
\label{sec21_14tris}
\left( \frac{\partial B}{\partial \phi^{(+)}} \right)^2 - \left( \frac{\partial B}{\partial \phi^{(-)}} \right)^2 = 2 (V^{(+)} - V^{(-)} )
\end{equation}
whereas expressions \eqref{sec2_14bis} solve the constraints  
\begin{subequations}
\label{sec21_11.1}
\begin{align}
  \label{eq:a}
  (1+E)(1-E)&=DD^T \\
  \label{eq:b}
 (1+F)(1-F)&=D^TD \\
  \label{eq:c}
 DF&=ED
\end{align}
\end{subequations}
and 
\begin{subequations}
\label{sec21_15}
\begin{align}
\label{eq:sec21_15a}
&D+D^T=2 \\
\label{eq:sec21_15b}
&\alpha_i D^T=-\alpha_{i+1}D\\
\label{eq:sec21_15c}
&\alpha_iD\alpha_j= 2 (\delta_{ij} - \delta_{i, j+1}) 
\end{align}
\end{subequations}
Conditions (\ref{sec21_14tris}--\ref{sec21_15}) arise by imposing that the defect potential does not spoil the conservation laws \cite{b18,b21}.  Again, these constraints ensure also momentum conservation, despite the lack of translational invariance of the model. Therefore, for $A_r^{(1)}$ Toda theories forcing momentum conservation seems to be a sufficient condition for ensuring the classical integrability of the theory. 
  
It is important to observe that \eqref{sec2_14bis} is only one of the infinite possible solutions to these constraints. In fact, suppose we act separately with two orthogonal transformations on the fields on the left and on the right of the defect
\begin{equation}
\phi^{(-)}\rightarrow O_1 \phi^{(-)} \hspace{3mm}, \hspace{3mm} \phi^{(+)}\rightarrow O_2 \phi^{(+)}
\end{equation}
These transformations leave the kinetic terms unchanged, while rotate the simple roots in the bulk potentials (\ref{intro1}) without modifying their scalar product. In the defect lagrangian (\ref{intro3})  they modify the defect matrices as
\begin{equation}
E\rightarrow O_1^T E O_1 \hspace{3mm},\hspace{3mm} F\rightarrow O_2^T F O_2 \hspace{3mm},\hspace{3mm} D\rightarrow O_1^T D O_2 
\label{sec21_11.2}
\end{equation}
Since these trasformations leave constraints (\ref{sec21_11.1}) invariant, they can be used to generate an infinite set of solutions starting from a given $D,E,F$ solution. In particular, each solution is connected to the one, $E=F=1-D$, through a pair of orthogonal transformations whose only effect is to rotate the left and right root bases differently.  

In general, while in the full-line Toda theory two conserved charges $Q_{\pm s}$ exist for a given spin-$(s+1)$ conserved current, in the presence of type-I defects only a linear combination of the two survives \cite{b18}. The particular linear combination depends on the solution to constraints \eqref{sec21_11.1} that we choose for the defect matrices. Applying an orthogonal transformation \eqref{sec21_11.2} to map one solution into another one, will then modify the expression of the corresponding conserved charges, however without spoiling the integrability properties of the theory.  

For $A_{r>1}^{(1)}$ theories this happens already for the spin-$(\pm 2)$ charges. In fact, as reviewed in details in appendix \ref{App:spin3}, in this case starting with equal root bases for the left and right theories, the solution $E=F=1-D$ leads to the cancellation of unwanted terms in the conservation laws of $Q_{+2}$, while breaking the conservation of $Q_{-2}$. Alternatively, the solution $E=F=-1-D$ guarantees the conservation of $Q_{-2}$ but not of $Q_{+2}$. 
These two possible choices for the defect matrices are related by a transformation of the form (\ref{sec21_11.2}). Each of them is an eligible solution and leads to a defect potential that preserves also energy and momentum \cite{b18}, although the two defect potentials are obviously different. 

To conclude this section it is worth observing that in all the theories, sinh-Gordon included,  there is a further choice for an integrable defect. This corresponds to set $D=E=F=0$ and choose $B(\phi^{(+)}, \phi^{(-)}) = B(\phi^{(+)}) + B(\phi^{(-)})$ with 
\begin{equation} \label{trivialcase}
B(\phi) = \sum_{j=0}^r d_j e^{\frac12 \alpha \cdot \phi} , \qquad \quad d_j^2 = 4
\end{equation} 
In this case the theory reduces to a double copy of two non-mutually interacting Toda theories with boundary, for which potential \eqref{trivialcase} is known to ensure conservation of energy-like currents at classical level \cite{a29, a30, a31}, whereas quantum conservation requires  in general a finite renormalization of the $d_j$ coefficients \cite{a32,a33,a34}.

\section{Quantum conservation of energy and momentum }\label{EM}

For a generic $A_r^{(1)}$ Toda theory we begin by studying the conservation of the stress-energy tensor. 
Since the defect breaks translational invariance along the $x_1$ axis, physically one should expect that in general momentum is no longer conserved and only the energy, related to translational invariance along the defect, survives as a conserved charge. However, the classical analysis carried out in \cite{b17,b18} has revealed that for this class of Toda theories it is possible to select a particular set of defects that contribute with the right amount of momentum to ensure conservation of the total charge. In addition, momentum conservation surprisingly implies conservation of an infinite number of higher-spin currents \cite{b21}. Therefore, at classical level integrability-preserving defects can be selected by constraints based solely on energy-momentum conservation. We want to investigate if the same pattern occurs at quantum level. 

To this end we consider the spin-2 current whose light-cone bulk components at the left and the right of the defect are given by
\begin{equation}
J^{(\pm)} \equiv T^{(\pm)}_{xx}=\partial \phi^{(\pm)}_a \partial \phi^{(\pm)}_a \hspace{4 mm}, \hspace{4 mm} \tilde{J}^{(\pm)} \equiv T^{(\pm)}_{\bar{x}\bar{x}}=\bar{\partial} \phi^{(\pm)}_a \bar{\partial} \phi^{(\pm)}_a
\label{sec21_1}
\end{equation}

From now on we adopt the following conventions for the derivatives of the bulk and defect potentials 
\begin{equation}\begin{split}
& V^{(\pm)}_{a_1 a_2\dots a_p}\equiv \frac{\partial^p V^{(\pm)}}{\partial \phi_{a_1}^{(\pm)}\partial \phi_{a_2}^{(\pm)} \dots \partial \phi_{a_p}^{(\pm)}} \\ 
& B_{a_1 a_2\dots a_p}^{i_1 i_2 \dots i_p} \equiv \frac{\partial^p B}{\partial \phi_{a_1}^{(i_1)}\partial \phi_{a_2}^{(i_2)} \dots \partial \phi_{a_p}^{(i_p)}} \qquad \quad 
i_1,i_2, \dots, i_p = +,-
\label{conventions}
\end{split}\end{equation}
Moreover, in the intermediate steps of the calculations we omit the upper case $\pm$ for bulk currents whenever this does not cause ambiguities.

According to the general procedure described in the previous section, we begin by evaluating bulk conservation laws \eqref{sec2_5}. Expanding the interaction action at first order and Taylor expanding the $V$ potential, at both sides of the defect we have 
\begin{equation}\begin{split}
\bar{\partial} \Bigl \langle T_{xx}(x)\Bigr \rangle &=-\bar{\partial} \Bigl \langle \partial \phi_a(x) \partial \phi_a(x) \frac{1}{\beta^2} \int d^2\omega V_{b} \, \phi_b(\omega)\Bigr \rangle_0\\
& \quad - \bar{\partial}\Bigl \langle\partial \phi_a(x) \partial \phi_a(x) \frac{1}{\beta^2}\int d^2\omega \frac{1}{2}V_{cd} \, \phi_c(\omega) \phi_d(\omega)\Bigr \rangle_0 \\
& = -\frac{2}{\beta^2}\bar{\partial}\int d^2\omega  \, V_{a}(\omega) \, \partial \phi_a (x) \, \Bigl(-\frac{\beta^2}{4 \pi}\Bigr)\bar{\partial} \biggl[\frac{1}{(x-\omega)} +\frac{1}{(x-\bar{\omega})}\biggr]\\
& \quad  -\frac{1}{\beta^2}\int d^2\omega V_{aa} \Bigl(-\frac{\beta^2}{4\pi}\Bigr)^2\bar{\partial} \biggl[\frac{1}{(x-\omega)} +\frac{1}{(x-\bar{\omega})}\biggr]^2 
\label{sec21_2}
\end{split}\end{equation}
where we have used propagators \eqref{sec2_3} to perform field contractions. 
Now, defining $\gamma \equiv \frac{\beta^2}{2\pi}$ and using identity (\ref{sec2_8}) we obtain
\begin{equation}\begin{split}
\bar{\partial}\bigl \langle T_{xx}(x)\bigr \rangle &= \int d^2\omega \, V_a \,  \partial \phi_a(x) \, \delta^{(2)}(x-\omega)-\frac{\gamma}{4} \int d^2\omega \, V_{aa}  \, \partial_\omega  \delta^{(2)} (x-\omega)\\
&=\partial \Bigl(V(x)+\frac{\gamma}{4} V_{aa}(x)\Bigr)
\label{sec21_3}
\end{split}\end{equation}
which for the potential in expression (\ref{intro1}) ($V_{aa} = 2V$) and recalling eq. \eqref{sec2_5} leads to the following trace terms
\begin{equation}
\Theta^{(\pm)} \equiv T^{(\pm)}_{x\bar{x}}= T^{(\pm)}_{\bar{x}x}=-\Bigl( 1+ \frac{\gamma}{2} \Bigr) V^{(\pm)} 
\label{sec21_4}
\end{equation}
Therefore, conservation law \eqref{intro7} for the stress-energy tensors in the bulk is valid also at quantum level, with a non-trivial quantum correction to the traces as given in the above equation. Of course, setting $\gamma=0$ we are back to the well-known classical expressions. 

The next step is to study the conservation of energy and momentum at the defect. 
We first consider the simpler case of energy conservation. According to the general discussion of section 2, the energy flow at the defect is given by the left hand side of eq. \eqref{sec2_9}, which using definitions \eqref{eq:int_b} takes the explicit form 
\begin{equation}\begin{split}
&2 \Bigl \langle T_{10}^{(+)}-T_{10}^{(-)}\Bigr \rangle\Bigr|_{x_1=0} = i \Bigl \{ \Bigl \langle T_{xx}^{(+)}-T_{\bar{x}\bar{x}}^{(+)}\Bigr \rangle-\Bigl \langle T_{xx}^{(-)}-T_{\bar{x}\bar{x}}^{(-)}\Bigr \rangle \Bigr \}\Bigr|_{x_1=0}\\
&=i\Bigl \{ \Bigl \langle \partial \phi_a^{(+)} \partial \phi_a^{(+)} - \bar{\partial} \phi_a^{(+)} \bar{\partial} \phi_a^{(+)}\Bigr \rangle-\Bigl \langle \partial \phi_a^{(-)} \partial \phi_a^{(-)} -\bar{\partial} \phi_a^{(-)} \bar{\partial} \phi_a^{(-)}\Bigr \rangle \Bigr \}\Bigr|_{x_1=0} 
\label{sec21_5}
\end{split}\end{equation}
We  evaluate the two terms of the second line separately. We report the explicit calculation for the first term, being the second term treated exactly in the same way. 

In this case local terms come only from contractions with the defect action $S_d = \int V^{(d)}$ with the $V^{(d)}$ potential given in \eqref{intro3}. Keeping only terms that actually contribute, we obtain
\begin{equation}\begin{split}
&\Bigl \langle \partial \phi^{(+)}_a \partial \phi^{(+)}_a-\bar{\partial} \phi^{(+)}_a \bar{\partial} \phi_a^{(+)} \Bigr \rangle\Bigl|_{x_1=0} \equiv \Bigl \langle \left( \partial \phi^{(+)}_a \partial \phi^{(+)}_a-\bar{\partial} \phi^{(+)}_a \bar{\partial} \phi_a^{(+)} \right) e^{\int V^{(d)}} \Bigr \rangle_0\Bigl|_{x_1=0}\\
&=\frac{1}{\beta^2}\int d\omega_0\Bigl \langle \Bigl( \partial \phi^{(+)}_a \partial \phi^{(+)}_a-\bar{\partial} \phi^{(+)}_a \bar{\partial} \phi_a^{(+)} \Bigr)(x) \\
&\qquad \qquad \times \Bigl( \frac{i}{2} \phi_b^{(+)} F_{bc} \partial_0 \phi_c^{(+)} +i \phi_b^{(-)} D_{bc} \partial_0 \phi_c^{(+)} + B  \Bigr)(\omega_0)\Bigr \rangle_0\Bigl|_{x_1=0}\\
&=\lim_{x_1\to0^+}  \frac{1}{\sqrt{2}\pi}  \int d\omega_0  \biggl(\frac{1}{(x-\omega)}-\frac{1}{(\bar{x}-\omega)} \biggr) \partial_0\phi_a^{(+)}(x) \biggl[-i F_{ac} \partial_0\phi_c^{(+)}+ i \partial_0 \phi_b^{(-)} D_{ba} -B_a^{+}\biggr](\omega)\\
&+\lim_{x_1\to0^+}\frac{\beta^2}{4\pi^2} \int d\omega_0 B_{aa}^{++}\biggl(\frac{1}{(x-\omega)^2}-\frac{1}{(\bar{x}-\omega)^2} \biggr) \\
& \\
&=2\partial_0 \phi^{(-)}_a D_{ab} \, \partial_0 \phi^{(+)}_b+2i\partial_0\phi_a^{(+)}B_a^{+}+2i \gamma \,\partial_0B_{aa}^{++}
\label{sec21_6}
\end{split}\end{equation}
Here free equations of motion at the defect, $\partial_1 \phi^{(+)}_a|_{x_1^+=0} = 0$, have been used to set $\partial \phi^{(+)}|=\bar{\partial} \phi^{(+)}|= \frac{1}{\sqrt{2}}\partial_0 \phi^{(+)}$.  Moreover, the last equality has been obtained using identity \eqref{sec2_12}.

Similarly, for the second term in eq. \eqref{sec21_5} we obtain
\begin{equation}
\Bigl \langle \partial \phi^{(-)}_a \partial \phi^{(-)}_a-\bar{\partial} \phi^{(-)}_a \bar{\partial} \phi_a^{(-)} \Bigr \rangle\Bigl|_{x_1=0} =2\partial_0 \phi^{(-)}_a D_{ab} \, \partial_0\phi^{(+)}_b-2i\partial_0\phi_a^{(-)}B_a^{-}-2i \gamma \, \partial_0B_{aa}^{--}
\label{sec21_7}
\end{equation}
Therefore, subtracting these two expressions as in (\ref{sec21_5}) we obtain the energy contribution carried by the defect
\begin{equation}
2 \Bigl \langle T_{10}^{(+)}-T_{10}^{(-)}\Bigr \rangle\Bigr|_{x_1=0} = - \partial_0\biggl(2B+2\gamma \, \bigl(B_{aa}^{++}+B_{aa}^{--}\bigr)\biggr) \equiv  2 \partial_0 \Sigma_E
\label{sec21_8}
\end{equation}
The total conserved energy is then given by $E^{(-)} + E^{(+)} - \Sigma_E$, where $E^{(\pm)}$ are the energies of the bulk systems. 

We note that in this case quantum local contributions give rise to non-trivial corrections to the currents and the corresponding charges, without spoiling the conservation law. Therefore, energy is always conserved for type-I defects, independently of the particular choice of the $D,E$ and $F$ matrices (with the only condition that $E$ and $F$ are antisymmetric) and the $B$ potential in \eqref{intro3}. We stress that the quantum corrections we have found are exact in $\gamma$, since higher order terms in the expansion of the actions in \eqref{sec21_2} and \eqref{sec21_6} would give non-local terms not ascribable to a conservation law. 

Proceeding in a similar way it is possible to compute quantum corrections to the momentum flow. This time we need to evaluate
\begin{equation}\begin{split} \label{sec21_5bis}
&-2i\Bigl \langle T_{11}^{(+)}-T_{11}^{(-)}\Bigr \rangle\Bigr|_{x_1=0} = i \Bigl \{ \Bigl \langle T_{xx}^{(+)}+T_{\bar{x}\bar{x}}^{(+)} - 2 T_{x \bar{x}}^{(+)} \Bigr \rangle - \Bigl \langle T_{xx}^{(-)}+T_{\bar{x}\bar{x}}^{(-)} - 2 T_{x \bar{x}}^{(-)}\Bigr \rangle \Bigr \}\Bigr|_{x_1=0}\\
&=i\Bigl \{ \Bigl \langle \partial \phi_a^{(+)} \partial \phi_a^{(+)} + \bar{\partial} \phi_a^{(+)} \bar{\partial} \phi_a^{(+)} \Bigr \rangle-\Bigl \langle \partial \phi_a^{(-)} \partial \phi_a^{(-)} + \bar{\partial} \phi_a^{(-)} \bar{\partial} \phi_a^{(-)}  \Bigr \rangle \Bigr \}\Bigr|_{x_1=0} \\
& \quad + 2i \left( 1 + \frac{\gamma}{2} \right) \Bigl \langle V^{(+)} - V^{(-)} \Bigr \rangle\Bigr|_{x_1=0}
\end{split}\end{equation}
We can proceed as done for the energy conservation. However, in this case the first order expansion in $S_d$ does not produce local contributions, since we would end up with the sum of inverse powers of $(z-\omega)$ instead of the difference and identity  \eqref{sec2_12} could not be used. Local contributions arise instead at zero and second order in the $S_d$-expansion. We obtain the zero order contribution simply by setting  $\partial_1 \phi^{(\pm)}=0$ in the current at the defect and performing no contractions. At second order, local terms come from performing two Wick contractions with $S_d^2 = \int d \omega_0 d \tilde{\omega}_0 V^{(d)}(\omega_0) V^{(d)}(\tilde{\omega}_0)$, so to remove the two integrals. The repeated use of identity \eqref{sec2_12} allows to obtain (see appendix \ref{App:momentumflow} for details)
\begin{equation}\begin{split}
& -2i\Bigl \langle T_{11}^{(+)}-T_{11}^{(-)}\Bigr \rangle\Bigr|_{x_1=0} = i\biggl[ \partial_0\phi^{(+)} \Bigl(1+F^TF-D^TD\Bigr)\partial_0\phi^{(+)}\\
&+\partial_0\phi^{(-)}\Bigl(-1-E^TE+DD^T\Bigr)\partial_0\phi^{(-)}+\partial_0\phi^{(-)}  \Bigl(-2DF-2E^TD\Bigr) \partial_0\phi^{(+)}\\
&+\partial_0\phi^{(-)}\Bigl(2iE^T B^- + 2i D B^+ \Bigr) +\partial_0\phi^{(+)}\Bigl(-2iF^T B^+ + 2i D^T B^- \Bigr)\\
&-(B^+)^2+(B^-)^2+2\Bigl(1+\frac{\gamma}{2}\Bigr) \Bigl( V^{(+)}-V^{(-)}\Bigr) \biggr] 
\label{sec21_11}
\end{split}\end{equation}
We note that no further powers of the coupling constant arise from these contractions, thus the result is identical to the classical one apart from the renormalization factor $\Bigl(1+\frac{\gamma}{2}\Bigr)$ in front of the trace.

Requiring this expression to be equal to $-2i \,\partial_0 \Sigma_P$ is non-trivial and leads to strong constraints on the form of $E$, $F$, $D$ matrices, as listed in (\ref{sec21_11.1}--\ref{sec21_15}). Moreover, taking into account the non-trivial renormalization of the trace, condition \eqref{sec21_14tris} gets modified by a factor $\Bigl(1+\frac{\gamma}{2}\Bigr)$ in from of the r.h.s. and the solution to this new constraint is 
\begin{equation}
B=\sqrt{1+\frac{\gamma}{2}}\, B^{(0)}
\label{sec21_13}
\end{equation}
where we have called $B^{(0)}$ the classical potential in \eqref{sec21_14}. 

Summarising, in order to preserve energy-momentum conservation at quantum level it is sufficient to renormalize the defect potential $B$ as in \eqref{sec21_13}, while maintaining the classical solution for the $D, E,F$ matrices. Exploiting the explicit expression of $B$ we can determine the exact values of  $\Sigma_E$ and $\Sigma_P$. If we define the following quantities
\begin{equation}
P^{(i)}_+=\frac{1}{\sigma}e^{\frac{1}{2}\alpha_i \cdot (D \phi^{(+)}+D^T\phi^{(-)})} \hspace{3mm}, \hspace{3mm} P^{(i)}_-=\sigma e^{\frac{1}{2}\alpha_i \cdot D(\phi^{(-)}-\phi^{(+)})}
\end{equation}
the contributions to energy and momentum carried by the defect are given by
\begin{subequations}
\label{sec21_16}
\begin{align}
  \label{eq:sec21_16a}
  \Sigma_E =& -\sqrt{1+\frac{\gamma}{2}}\sum_{i=0}^r \Bigl( 1+\frac{\gamma}{2} \alpha_i D D^T \alpha_i \Bigr) \Bigl( P^{(i)}_+ + P^{(i)}_- \Bigr) \\
  \label{eq:sec21_16b}
 \Sigma_P =& -i\sqrt{1+\frac{\gamma}{2}}\sum_{i=0}^r\Bigl( P^{(i)}_+ - P^{(i)}_- \Bigr)
\end{align}
\end{subequations}
At classical level ($\gamma=0$) these expressions acquire a more symmetric form and can be thought of as the complex components of a spin-1 charge defined on the defect. However, quantum corrections introduce an asymmetry in the $\gamma$-dependence and the nice interpretation in terms of complex components gets spoiled, suggesting that things might get worse when studying higher-spin currents. \\

\section{Anomalies in higher-spin currents}\label{higherspin}

We now move to the study of higher-spin conservation laws in $A_r^{(1)}$ Toda theories. At classical level it has been proved  \cite{b18,b21} that the same defect potential conserving the momentum flow guarantees the conservation of an infinite tower of higher spin charges, so ensuring integrability. In order to investigate whether the same pattern remains true at quantum level we study the first non-trivial conserved current with spin $s+1>2$. 
For sinh-Gordon model this means spin-4 current, whereas for higher rank Toda theories this corresponds to spin-3 current. We study the two cases separately.

\subsection{Spin-4 current in sinh-Gordon model}

As reviewed in section \ref{classical}, for the sinh-Gordon or $A^{(1)}_1$ Toda theory the bulk potential is given in \eqref{sec31_1}, whereas on the defect we choose 
\begin{equation}
V^{(d)} = i D \,\phi^{(-)}\partial_{0}\phi^{(+)}+B(\phi^{(-)},\phi^{(+)})
\label{sec31_2}
\end{equation}
where $D$ is a real number. According to the computation performed in the previous section momentum is preserved at the defect if we take $D=1$ and $B = \sqrt{1 + \tfrac{\gamma}{2}} B^{(0)}$, with $B^{(0)}$ given in \eqref{sec31_4}.  As already stressed, classically these conditions are sufficient to ensure an infinite number of conservations laws, the first being a spin-4 current. We here study the quantum counterpart of this current. 
 
The most general structure for the $J$-component of a spin-4 current on both sides of the defect is (we neglect $\pm$ labels)
\begin{equation}
J=\frac{a}{4}(\partial\phi)^4 + \frac{d}{2} (\partial^2 \phi)^2 
\label{sec31_5}
\end{equation}
where, for the moment, we omit total $\partial$-derivative terms.

Using the procedure explained in section \ref{procedure} we first study the bulk conservation law. Computing $\bar{\partial}\langle J  \rangle$ at both sides of the defect we find
\begin{equation}\begin{split}
\bar{\partial}\Bigl \langle J \Bigr \rangle&=\partial \biggl[ \left( \frac{d}{4} \frac{\partial^2 V}{\partial \phi^2} + \gamma^2 \frac{a}{32} \frac{\partial^4 V}{\partial \phi^4}  \right) (\partial \phi)^2   + \gamma \frac{d}{48} \partial^2 \frac{\partial^2 V}{\partial \phi^2} +\gamma^3 \, \frac{a}{384} \partial^2 \frac{\partial^4 V}{\partial \phi^4} \biggr]\\
&+\frac{1}{2}\biggl[a \frac{\partial V }{\partial \phi }+(-\frac{d}{2}+\gamma \, \frac{3}{4} a) \frac{\partial^3 V }{\partial \phi^3} + \gamma^2 \frac{a}{16} \frac{\partial^5 V }{\partial \phi^5}\biggl] (\partial \phi )^3 
\label{sec31_6}
\end{split}\end{equation}
The bulk anomalous contribution in the second line can be cancelled by imposing
\begin{equation}
a \frac{\partial V}{\partial \phi}+(-\frac{d}{2}+\gamma \frac{3}{4} a) \frac{\partial^3 V}{\partial \phi^3} + \gamma^2 \frac{a}{16} \frac{\partial^5 V}{\partial \phi^5}=0
\label{sec31_7}
\end{equation}
Given that the potential in (\ref{sec31_1}) satisfies $\frac{\partial^3 V}{\partial \phi^3} = 2 \frac{\partial V}{\partial \phi}$, $\frac{\partial^5 V}{\partial \phi^5} = 4 \frac{\partial V}{\partial \phi}$, the solution reads
\begin{equation}
d=a\Bigl(1+\frac{3}{2}\gamma + \frac{\gamma^2}{4}\Bigr) 
\label{sec31_8}
\end{equation}
At the same time, recalling the general structure \eqref{intro7} for a conservation law in complex coordinates, from the first line of eq. (\ref{sec31_6}) we can read the trace $\Theta$. Therefore, using solution \eqref{sec31_8}, up to an overall normalization factor the quantum bulk currents are given by
\begin{equation}\begin{split}
& J^{(\pm)}= (\partial\phi^{(\pm)})^4 + \Bigl(2+ 3\gamma+ \frac{\gamma^2}{2}\Bigr)  (\partial^2 \phi^{(\pm)})^2  \\
& \Theta^{(\pm)}= - \biggl(1+ \frac52 \gamma + \frac34 \gamma^2 + \frac{1}{12}\gamma^3 \biggr) \frac{\partial^2 V^{(\pm)}}{\partial {\phi^{(\pm)}}^2} (\partial \phi^{(\pm)})^2
 -\frac{\gamma}{24} \, \biggl(4+6 \gamma + 2\gamma^2 \biggr) \frac{\partial V^{(\pm)}}{\partial \phi^{(i)}}\partial^2 \phi^{(\pm)}
\label{sec31_9}
\end{split}\end{equation}
The expressions for $\tilde{J}^{(\pm)}$ and $\tilde{ \Theta}^{(\pm)}$ can be obtained by simply replacing $\partial \to \bar{\partial}$. 
Setting $\gamma = 0$ we are back to the classical expressions for the spin-4 bulk currents \cite{b17}. 

The second step is to evaluate the flow of the energy-like and momentum-like currents through the defect. Starting from the energy-like current, we evaluate 
\begin{equation}\begin{split}
&\Bigl \langle J^{(+)}_1-J^{(-)}_1 \Bigr \rangle \Bigr|_{x_1=0}=\\
&i\biggl[ \Bigl \langle J^{(+)}-\tilde{J}^{(+)}\Bigr \rangle - \Bigl \langle J^{(-)}-\tilde{J}^{(-)} \Bigr \rangle+\Bigl \langle \Theta^{(-)}-\tilde{\Theta}^{(-)}\Bigr \rangle -\Bigl \langle \Theta^{(+)}-\tilde{\Theta}^{(+)}\Bigr \rangle \biggr]\bigg|_{x_1=0} 
\label{sec31_10}
\end{split}\end{equation}
and impose it to be a total $\partial_0$-derivative.

We use the general method described in section \ref{procedure} and already highlighted in the case of energy and momentum conservation. 
Since the calculation is quite cumbersome, we organize the discussion by considering contributions with different powers of $\partial_0$-derivatives, separately. 

First of all, in the total result terms proportional to higher-order $\partial_0$-derivatives appear, which have the form
\begin{equation}\label{sec51_1}
\Bigl \langle J^{(+)}_1-J^{(-)}_1 \Bigr \rangle \Bigr|_{x_1=0} \longrightarrow  B^{---+} \, \partial_0^2 \phi^{(-)} \partial_0 \phi^{(+)}+ B^{+++-} \, \partial_0^2 \phi^{(+)} \partial_0 \phi^{(-)} 
\end{equation}
where we have used notation \eqref{conventions} for derivatives of $B$ respect to the $\phi^{(\pm)}$ fields. 
These contributions come from performing three contractions of the terms $[(\partial \phi^{(\pm)})^4 - (\bar{\partial} \phi^{(\pm)})^4]$ in \eqref{sec31_10}  with the defect potential $B$. The only possibility to reduce them to a total derivative is to integrate by parts $\partial_0$ in one of the two terms and require the $B$ potential to satisfy the constraint
\begin{equation}
B^{+++-} = B^{---+}
\label{sec51_3}
\end{equation}

The rest of the contributions can be reduced to powers of first order $\partial_0$-derivatives of the fields, up to total-derivative terms, and can be organized on the basis of the number of them. Using condition \eqref{sec51_3} and keeping the $D$ coefficient in \eqref{sec31_2} generic, forth order contributions in the field derivatives come only from $\langle J^{(\pm)} - \tilde{J}^{(\pm)} \rangle$ and read 
\begin{equation}\begin{split} \label{sec51_1bis}
&\Bigl \langle J_1^{(+)}-J_1^{(-)}\Bigr \rangle \Bigl|_{x_1=0}^{(4)} = 2i (D^3-D) (\partial_0\phi^{(-)})^3\partial_0\phi^{(+)}  - 2i (D^3-D)(\partial_0\phi^{(+)})^3\partial_0\phi^{(-)}
\end{split}\end{equation}
Cubic order terms in the field derivatives, up to total $\partial_0$-derivatives are 
\begin{equation}\begin{split}\label{sec51_2}
&\Bigl \langle J_1^{(+)}-J_1^{(-)}\Bigr \rangle \Bigl|_{x_1=0}^{(3)} =\\
& - \Bigl[ 2 B^+ - 4B^{+++} + 6 \gamma D^2 B^{--+} +2\gamma^2 B^{+++++} -\gamma^2 B^{+++} \Bigr] (\partial_0\phi^{(+)})^3\\
&- \Bigl[ 2 B^- - 4 B^{---} + 6 \gamma D^2 B^{++-} +2 \gamma^2 B^{-----}- \gamma^2 B^{---} \Bigr] (\partial_0\phi^{(-)})^3\\
& - \Bigl[ 6 D^2 B^+ -\left( 12  + 12\gamma + 3 \gamma^2 \right) B^{--+}   + 6 \gamma D^2 B^{+++} +6  \gamma^2 B^{+++--} \Bigr] (\partial_0\phi^{(-)})^2 \partial_0\phi^{(+)}\\
& - \Bigl[ 6 D^2 B^- -\left( 12  + 12\gamma + 3 \gamma^2 \right) B^{++-}   + 6\gamma D^2 B^{---} +6 \gamma^2 B^{---++} \Bigr] (\partial_0\phi^{(+)})^2 \partial_0\phi^{(-)} 
\end{split}\end{equation}
Finally, terms quadratic and linear in the field derivatives are
\begin{equation}\begin{split}\label{sec51_2bis}
&\Bigl \langle J_1^{(+)}-J_1^{(-)}\Bigr \rangle \Bigl|_{x_1=0}^{(2)} = \; 12 i \gamma D B^{+--} B^- (\partial_0\phi^{(+)})^2-12 i  \gamma D B^{++-} B^+  (\partial_0\phi^{(-)})^2\\
&-iD\Bigl[6 \Bigl(\bigl(B^{+}\bigr)^2-\bigl(  B^{-}\bigr)^2\Bigr) + 12 \gamma \Bigl(B^+ B^{+++} - B^- B^{---}\Bigr) \\
&\qquad \quad -\bigl(6+ 9\gamma + 2\gamma^2 \bigr) \Bigl( \frac{\partial^2 V^{(+)}}{{\partial \phi^{(+)}}^2}-\frac{\partial^2 V^{(-)}}{{\partial \phi^{(-)}}^2} \Bigr) \Bigr] \partial_0\phi^{(-)} \partial_0\phi^{(+)}
\end{split}\end{equation}
and
\begin{equation}\begin{split}\label{sec51_2bisbis}
&\Bigl \langle J_1^{(+)}-J_1^{(-)}\Bigr \rangle \Bigl|_{x_1=0}^{(1)}=\\
& +\Bigl[ 2 (B^{+})^3 +6 \gamma \Bigl((B^{+})^2B^{+++}+(B^{-})^2B^{--+}\Bigr)\\
& \quad + \gamma \Bigl(4+\frac{20}{3} \gamma +3 \gamma^2 + \frac{\gamma^3}{3}\Bigr) \frac{\partial V^{(+)}}{\partial \phi^{(+)}} B^{++} -\Bigl(6+9 \gamma+ 2 \gamma^2 \Bigr) \frac{\partial^2 V^{(+)}}{{\partial \phi^{(+)}}^2} B^{+}\Bigr] \, \partial_0 \phi^{(+)} \\
&+  \Bigl[ 2 (B^{-})^3 +6 \gamma \Bigl((B^{+})^2B^{++-}+(B^{-})^2B^{---}\Bigr)\\
& \quad + \gamma \Bigl(4+\frac{20}{3} \gamma +3 \gamma^2 + \frac{\gamma^3}{3}\Bigr) \frac{\partial V^{(-)}}{\partial \phi^{(-)}} B^{--} -\Bigl(6 +9 \gamma+2\gamma^2 \Bigr) \frac{\partial^2 V^{(-)}}{{\partial \phi^{(-)}}^2} B^{-}\Bigr] \, \partial_0 \phi^{(-)}
\end{split}\end{equation}

We begin by considering terms \eqref{sec51_1bis}. Since these contributions do not have any chance to reduce to a total derivative, they have to cancel. This amounts to require $D^3=D$, whose solutions are $D=0$ and $D=\pm1$. While the non-trivial solutions ensure momentum conservation, the trivial one breaks it already at classical level (see discussion in section \ref{EM}). Nonetheless, under a suitable choice of the defect potential it might still preserve higher-spin currents. Thus, we discuss the two cases separately. 

\vskip 10pt
\noindent
\underline{$D=0$}

\noindent
Setting $D=0$ reduces the defect potential to $V^{(d)} = B$. From the previous expressions we immediately see that in this case eqs. (\ref{sec51_1bis}, \ref{sec51_2bis}) vanish identically, whereas \eqref{sec51_2} becomes 
\begin{equation}\begin{split}\label{sec51_4}
&\Bigl \langle J_1^{(+)}-J_1^{(-)}\Bigr \rangle \Bigl|_{x_1=0}^{(3)} = \\
&+\Bigl[- 2 B^+ + \left( 4 + \gamma^2  \right) B^{+++} -2 \gamma^2 B^{+++++}\Bigr](\partial_0\phi^{(+)})^3\\
&+ \Bigl[-2 B^- + \left( 4 + \gamma^2  \right) B^{---} - 2 \gamma^2 B^{-----}\Bigr](\partial_0\phi^{(-)})^3\\
&+ \Bigl[\left( 12 + 12\gamma + 3 \gamma^2 \right) B^{--+} -6 \gamma^2 B^{+++--} \Bigl](\partial_0\phi^{(-)})^2 \partial_0\phi^{(+)}\\
&+\Bigl[\left( 12 + 12\gamma + 3 \gamma^2 \right) B^{++-} -6 \gamma^2 B^{---++} \Bigl]   (\partial_0\phi^{(+)})^2 \partial_0\phi^{(-)}
\end{split}\end{equation}
The first two lines of this expression are zero if we impose the further constraint 
\begin{equation}\label{sec51_5}
B^{+++}=\frac{1}{2} B^+ \hspace{3mm},\hspace{3mm} B^{---} =\frac{1}{2} B^-
\end{equation}
Now using these equations in the last two lines we are left with
\begin{equation}\begin{split}\label{sect51_6}
\Bigl \langle J_1^{(+)}-J_1^{(-)}\Bigr \rangle \Bigl|_{x_1=0}^{(3)} &=   12 B^{--+} \Bigl(1+  \gamma   \Bigr) (\partial_0\phi^{(-)})^2 \partial_0\phi^{(+)} \\
&+ 12 B^{++-} \Bigl( 1+  \gamma \Bigr)  (\partial_0\phi^{(+)})^2 \partial_0\phi^{(-)}
\end{split}\end{equation}
which eventually vanishes if 
\begin{equation}
B^{--+} = B^{++-} =0
\label{sec51_6}
\end{equation}
Collecting contraints (\ref{sec51_3}, \ref{sec51_5}, \ref{sec51_6}) the most general solution for $B$ is  
\begin{equation}\label{sec51_7}
B=\Bigr(d^{(-)}_0 e^{-\frac{1}{\sqrt{2}}\phi^{(-)}}+d^{(-)}_1 e^{\frac{1}{\sqrt{2}}\phi^{(-)}} \Bigl)\, +\, \Bigr(d^{(+)}_0 e^{-\frac{1}{\sqrt{2}}\phi^{(+)}}+d^{(+)}_1 e^{\frac{1}{\sqrt{2}}\phi^{(+)}}\Bigl)
\end{equation}
and the model reduces to two decoupled theories on the left and on the right, with non-trivial but independent boundaries potentials. With this form of the potential the linear terms in \eqref{sec51_2bisbis} take the  form $f(\phi^{(+)}) \partial_0 \phi^{(+)}+g(\phi^{(-)}) \partial_0 \phi^{(-)}$ which is always a total $\partial_0$-derivative, as long as $f$ and $g$ admit a primitive. This case corresponds simply to two copies of  the sinh-Gordon model with boundary studied in \cite{a32,a33,a34}.  In particular, defect potential \eqref{sec51_7} ensures current conservation both at classical and quantum level, without any finite renormalization of its coefficients.

\vskip 10pt
\noindent
\underline{$D=1$}

\noindent
We now consider the two solutions $D=\pm1$, which cancel contributions \eqref{sec51_1bis}. Since the transformation $D \to -D$ is equivalent to exchange $\phi^{(-)} \leftrightarrow \phi^{(+)}$ in the defect lagrangian it is not restrictive to choose $D=1$. As discussed in section \ref{EM},  this is exactly one of the conditions that ensures also momentum conservation.

Setting $D=1$ in eqs. (\ref{sec51_2} - \ref{sec51_2bisbis}) we obtain 
\begin{equation}\begin{split}\label{sec51_8}
&\Bigl \langle J_1^{(+)}-J_1^{(-)}\Bigr \rangle \Bigl|_{x_1=0}^{(3)}=\\
&- 2\Bigl[B^+ -\left( 2 + \frac{\gamma^2}{2} \right) B^{+++} +3 \gamma  B^{--+}+\gamma^2 B^{+++++} \Bigr](\partial_0\phi^{(+)})^3\\
&- 2\Bigl[B^- -\left( 2 + \frac{\gamma^2}{2} \right) B^{---} +3 \gamma  B^{++-}+\gamma^2 B^{-----} \Bigr] (\partial_0\phi^{(-)})^3\\
& - 6\Bigl[ B^+ -\left( 2 + 2 \gamma + \frac{\gamma^2}{2} \right) B^{--+}+  \gamma  B^{+++}+ \gamma^2 B^{+++--} \Bigr](\partial_0\phi^{(-)})^2 \partial_0\phi^{(+)} \\
&- 6 \Bigl[ B^- -\left( 2 + 2 \gamma + \frac{\gamma^2}{2} \right) B^{++-}+  \gamma  B^{---}+ \gamma^2 B^{---++} \Bigr] (\partial_0\phi^{(+)})^2 \partial_0\phi^{(-)} 
\end{split}\end{equation}
and
\begin{equation}\begin{split}\label{sec51_9}
&\Bigl \langle J_1^{(+)}-J_1^{(-)}\Bigr \rangle \Bigl|_{x_1=0}^{(2)}= \; 12i \gamma  B^{--+} B^-(\partial_0\phi^{(+)})^2-12i  \gamma B^{++-} B^+ (\partial_0\phi^{(-)})^2\\
&-i\Bigl[6 \Bigl( B^+ \bigl)^2-\bigr( B^- \bigl)^2\Bigr) + 12 \gamma \Bigl(B^+B^{+++} - B^-B^{---} \Bigr) \\
&\qquad -\bigl(6+ 9\gamma +2 \gamma^2 \bigr) \Bigl( \frac{\partial^2 V^{(+)}}{{\partial \phi^{(+)}}^2}-\frac{\partial^2 V^{(-)}}{{\partial \phi^{(-)}}^2} \Bigr) \Bigr] \partial_0\phi^{(-)} \partial_0\phi^{(+)}
\end{split}\end{equation}
In the limit $\gamma \to 0$, using the classical defect potential in (\ref{sec31_4}) determined by imposing momentum conservation, these quantities indeed vanish. This is consistent with the classical findings and indicates that the momentum conservation is a good probe to verify classical integrability, as already discussed \cite{b18,b21}. 

However, at quantum level the situation becomes much worse. In fact, due to $\gamma$-corrections, we should require for instance that the terms proportional to $(\partial_0\phi^{(-)})^2$ and $(\partial_0\phi^{(+)})^2$  in (\ref{sec51_9}) both vanish. This necessarily implies that the cubic derivatives of $B$ have to vanish at each order in the $\gamma$-expansion, so leading to a defect potential that already at classical level differs from \eqref{sec31_4}. Therefore, due to irremovable anomalous terms that spoil eq. \eqref{sec31_10}, the spin-4 current is not conserved at the defect.

We might hope that, in analogy with the classical case, forcing momentum conservation would improve the situation. If we use renormalization condition (\ref{sec21_13}) to ensure momentum conservation, the expression \eqref{sec51_8} reduces to
\begin{equation}
\Bigl \langle J_1^{(+)}-J_1^{(-)}\Bigr \rangle \Bigl|_{x_1=0}^{(3)} = 3 \gamma \left( B^- \partial_0\phi^{(-)} - 
B^+\partial_0\phi^{(+)} \right) \left[ (\partial_0\phi^{(+)})^2 - (\partial_0\phi^{(-)})^2 \right]
\label{sec31_11bis}
\end{equation}
which still is not a total $\partial_0$-derivative.

Similar anomalies are found also at lower orders in the field derivatives, indicating that no renormalization of the defect potential exists, which can absorb these anomalous contributions. 

As discussed in section \ref{procedure}, a possible way-out could be the addition of total $\partial$-derivatives in the original structure \eqref{sec31_5} of the current.  We then discuss this possibility.  We start with the more general expression for the spin-4 currents
\begin{equation}
J^{(\pm)}=\frac{a}{4} (\partial \phi^{(\pm)})^4 + \frac{d}{2} (\partial^2\phi^{(\pm)})^2 + c \partial(\partial^2\phi^{(\pm)} \partial \phi^{(\pm)}) + e \partial ((\partial \phi^{(\pm)})^3)
\label{sec31_14}
\end{equation}
The conservation laws of the bulk currents now acquire extra terms
\begin{equation}\begin{split}
\bar{\partial} \Bigl \langle J^{(\pm)}\Bigr \rangle \rightarrow  \bar{\partial} \Bigl \langle J^{(\pm)}\Bigr \rangle &+ c \,\partial \biggl( \frac{1}{2} \partial \frac{\partial V^{(\pm)}}{\partial \phi^{(\pm)}} \partial \phi^{(\pm)}+\frac{1}{2} \frac{\partial V^{(\pm)}}{\partial \phi^{(\pm)}} \partial^2\phi^{(\pm)} + \frac{\gamma}{8}\partial^2 \frac{\partial^2V^{(\pm)}}{{\partial \phi^{(\pm)}}^2} \biggr)\\
&+ e \, \partial \biggl( \frac{3}{2} \frac{\partial V^{(\pm)}}{\partial \phi^{(\pm)}} (\partial \phi^{(\pm)})^2 + \frac{3}{4} \gamma \,\partial \phi^{(\pm)} \partial \frac{\partial^2V^{(\pm)}}{{\partial \phi^{(\pm)}}^2} + \frac{\gamma^2}{16}\partial^2\frac{\partial^3V^{(\pm)}}{{\partial \phi^{(\pm)}}^3}\biggr) 
\label{sec31_17}
\end{split}\end{equation}
From this expression we immediately find how the trace $\Theta$ gets modified. Substituting the new expressions for $J^{(\pm)}$, $\Theta^{(\pm)}$, $\tilde{J}^{(\pm)}$, $\tilde{\Theta}^{(\pm)}$ in the expectation value of $J_1^{(\pm)}$ we find that new terms appear. Without giving details of the calculation,
up to total $\partial_0$-derivatives the result reads 
\begin{equation}\begin{split}
 \Bigl \langle J_1^{(\pm)}\Bigr \rangle \Bigr|_{x_1=0} \rightarrow  \Bigl \langle J_1^{(\pm)}\Bigr \rangle \Bigr|_{x_1=0} & \mp c \biggl[ \gamma  \bigl(2+\gamma \bigr) \frac{\partial^2 B}{{\partial \phi^{(\pm)}}^2} \partial_0V^{(\pm)} \biggr]\biggr|_{x_1=0} + i e \biggl[ 3 \gamma \frac{\partial V^{(\pm)}}{\partial \phi^{(\pm)}} \partial_0 \phi^{(\pm)} \partial_0\phi^{(\mp)}\\
& \hspace{-2cm} \mp \frac{3}{2} i \gamma  \biggl(2 \frac{\partial V^{(\pm)}}{\partial \phi^{(\pm)}} \frac{\partial B}{\partial \phi^{(\pm)}}-4 \gamma V^{(\pm)} \frac{\partial^2 B}{{\partial \phi^{(\pm)}}^2}-\frac{2}{3}\gamma^2 V^{(\pm)} \frac{\partial^2 B}{{\partial \phi^{(\pm)}}^2} \biggr) \partial_0 \phi^{(\pm)}\biggr]\biggr|_{x_1=0} 
\label{sec31_20}
\end{split}\end{equation}
Unfortunately, the new $c$ and $e$-terms do not contain either cubic or quadratic powers of field derivatives that are required to cancel anomalous contributions in (\ref{sec51_8}, \ref{sec51_9}).  Therefore, we conclude that in the presence of defect \eqref{intro3} the energy-like spin-4 current is irretrievably anomalous at quantum level. 

The investigation of the momentum-like spin-4 current is then in order. This time, starting from the bulk currents \eqref{sec31_9} we need to evaluate $\Bigl \langle Y_1^{(+)}  -Y_1^{(-)}\Bigr \rangle \Bigl|_{x_1=0}$ where $Y_1^{(\pm)}$ are the linear combinations defined in \eqref{intro10}. Setting $D=1$ and $B^{++} = B^{--}$ (to cancel higher derivative terms), and neglecting total $\partial_0$-derivatives, after a quite long but straightforward calculation for the third-order contributions in the field derivatives we find
\begin{equation}\begin{split}
&\hspace{-0.2cm} \Bigl \langle Y_1^{(+)}  -Y_1^{(-)}\Bigr \rangle \Bigl|_{x_1=0}^{(3)} \equiv 
 i\biggl[ \Bigl \langle J^{(+)}+\tilde{J}^{(+)}\Bigr \rangle \hspace{-0.1cm}  - \hspace{-0.1cm} \Bigl \langle J^{(-)}+\tilde{J}^{(-)} \Bigr \rangle\hspace{-0.1cm} +\hspace{-0.1cm}\Bigl \langle \Theta^{(-)}+\tilde{\Theta}^{(-)}\Bigr \rangle \hspace{-0.1cm}-\hspace{-0.1cm}\Bigl \langle \Theta^{(+)} + \tilde{\Theta}^{(+)}\Bigr \rangle \biggr]\bigg|_{x_1=0}^{(3)}   \\
&= \; \; - 2\biggl[  B^- -\left( 2 + 3 \gamma + \frac{\gamma^2}{2} \right) B^{---}  + \gamma^2 B^{-----}  \biggr] (\partial_0\phi^{(+)})^3 \\
& \quad \; \; - 2\biggl[ B^+ -\left( 2 + 3 \gamma + \frac{\gamma^2}{2} \right) B^{+++}  +  \gamma^2 B^{+++++}  \biggr] (\partial_0\phi^{(-)})^3\\
&\quad -6 \biggl[  B^- -\left( 2 -  \gamma + \frac{\gamma^2}{2} \right) B^{---}  +\gamma^2 B^{-----}  \biggr] (\partial_0\phi^{(-)})^2 \partial_0\phi^{(+)}  \\
&\quad -6 \biggl[  B^+ -\left( 2 -  \gamma + \frac{\gamma^2}{2} \right) B^{+++}  +\gamma^2 B^{+++++}  \biggr]  (\partial_0\phi^{(+)})^2 \partial_0\phi^{(-)}
\label{sec31_11bisbis}
\end{split}\end{equation}
In this case it is impossible to require for instance that the first and third terms both vanish. The two terms, up to a multiplicative factor, differ for cubic derivatives of the defect potential, so they would be both zero only if these derivatives were to vanish,  but already at classical level this is clearly not the case. Therefore, also the momentum-like component of the current is lost at quantum level. 
Using the same kind of reasoning it can be proved that also any linear combination of energy-like and momentum-like currents is always anomalous, as long as the defect interacts non-trivially with the bulk fields.  

\vskip 10pt
There is, however, a rescuing situation. In the $D=1$ case, if we assume the defect potential to be multiplicatively renormalized, $B = f(\gamma) B^{(0)}$, with $B^{(0)}$ given in \eqref{sec31_4}, and imposing the extra constraint 
\begin{equation}
\partial_0 \phi^{(-)}\Bigr|_{x_1=0}=\partial_0 \phi^{(+)} \Bigr|_{x_1=0}
\label{sec31_11bisbisbis}
\end{equation}
then cubic order anomalies in \eqref{sec51_8} cancel, thanks to the identity $B^{++} = \frac12 B = B^{--}$, and terms \eqref{sec51_2bisbis} reduce automatically to a total $\partial_0$-derivative, since now $\phi^{(+)}$ and $\phi^{(-)}$ on the defect only differ by a constant. The only surviving terms come from \eqref{sec51_2bis} and they vanish if $B$ satisfies the following condition
\begin{equation} \label{sec51_11}
(3+3\gamma) \Bigl( (B^+)^2-(B^-)^2\Bigr)\Bigr) =2 \bigl(3+ \frac{9}{2}\gamma +\gamma^2\bigr)  \Bigl(  V^{(+)}- V^{(-)} \Bigr) 
\end{equation}
This implies the finite renormalization of the defect potential 
\begin{equation}\label{sec51_10}
B=(1 + \frac{\gamma}{4} -\frac{11}{96}\gamma^2+ \ldots )B^{(0)}
\end{equation}
which then ensures conservation of the energy-like current. 
It is important to stress that this renormalization is different from the one required for momentum conservation (see eq. (\ref{sec21_13})). We then conclude that in general at quantum level momentum conservation through the defect cannot be taken as a probe to verify the integrability of the theory. 

\vskip 10pt
Excluding this last case, spin-4 current and the corresponding spin-3 charge are spoiled by a non-trivial defect. Nevertheless, one might hope that an infinite number of higher-spin conserved charges inherited by the full-line sinh-Gordon theory still survive. Although we do not have a rigorous argument to prove the contrary, we expect the behaviour of the currents to become worse and worse as we increase their spin. In fact, higher and higher powers in the field derivatives arise and the number of anomalous terms that need to be cancel get out of control soon. This strongly leads to  suspect that at quantum level the sinh-Gordon model with a type-I defect of the form \eqref{intro3} looses its integrable properties. \\

\subsection{Spin-3 current in $A_r^{(1)}$ Toda models}\label{sect:spin3}

We now study the general case of $A_{r}^{(1)}$ Toda theories with $r>1$. For these models the first non-trivial conserved current beyond the stress-energy tensor is at spin 3.
Its most general expression at the right and left of the defect reads
\begin{equation}
J=\frac{1}{3}a_{abc}\partial \phi_a \partial \phi_b \partial \phi_c + b_{ab} \partial^2\phi_a \partial \phi_b +\frac{1}{2} c_{ab}\partial (\partial \phi_a \partial \phi_b)
\label{sec32_1}
\end{equation}
with $a_{abc}$, $c_{ab}$ totally symmetric and $b_{ab}$ antisymmetric constant parameters. This time we have already included a total derivative term since, as explained above, it can give rise to non-trivial quantum anomalies.

Studying the bulk conservation law along the lines described in section \ref{procedure} we obtain 
\begin{equation}\begin{split}
&\bar{\partial}\bigl \langle J \bigr \rangle= \partial \Bigl[\frac{1}{2} b_{ab} V_b\partial \phi_a + \frac{1}{2}c_{ab} V_b \, \partial \phi_a + \frac{\gamma}{8} c_{ab} \, \partial V_{ab} + \frac{\gamma^2}{48} a_{abc} \, \partial V_{abc}\Bigr]\\
&\qquad \quad +\frac{1}{2}\Bigl[a_{abc}V_a+b_{ac}V_{ab}+b_{ab}V_{ac}+\frac{\gamma}{4}a_{abd}V_{acd}+\frac{\gamma}{4}a_{acd}V_{abd}\Bigr]\partial \phi_b\partial \phi_c 
\label{sec32_2}
\end{split}\end{equation}
While the first term at the r.h.s. of this equation is already in the form of a total $\partial$-derivative, the second line needs to vanish if we want the bulk conservation to be still valid. Indeed, using  potentials $V^{(\pm)}$ in \eqref{intro1}, the second term gets cancelled if for each simple root $\alpha_i$ at both sides of the defect we impose the following constraint (no summation over $i=0, \dots , r$)
\begin{equation}
a_{abc}(\alpha_i)_c + b_{cb}(\alpha_i)_c(\alpha_i)_a+b_{ca}(\alpha_i)_c(\alpha_i)_b+\frac{\gamma}{4}a_{cbd}(\alpha_i)_c(\alpha_i)_a(\alpha_i)_d 
+\frac{\gamma}{4}a_{cad}(\alpha_i)_c(\alpha_i)_b(\alpha_i)_d=0
\label{sec32_3}
\end{equation}
whereas the matrix $c_{ab}$ is totally free. 

Calling $a_{abc}^{(0)}$ and $b_{ab}^{(0)}$ the classical parameters satisfying
\begin{equation}
a^{(0)}_{abc}(\alpha_i)_c + b^{(0)}_{cb}(\alpha_i)_c(\alpha_i)_a+b^{(0)}_{ca}(\alpha_i)_c(\alpha_i)_b=0
\label{sec32_4}
\end{equation}
relations (\ref{sec32_3}) hold for 
\begin{equation}
a_{abc}=a^{(0)}_{abc}  \hspace{4 mm},\hspace{4 mm} b_{ab}=\Bigl(1+\frac{\gamma}{2}\Bigr)b_{ab}^{(0)}
\label{sec32_5}
\end{equation}
Without loosing generality, up to an overall constant factor we can choose the $b^{(0)}$ coefficients to satisfy $b^{(0)} = 1 - D^{(0)}$, where $D^{(0)}$ is the classical solution \eqref{sec2_14bis} \cite{b18}. Plugging this expression in \eqref{sec32_4} leads to a consistent set of equations that allow to determine the $a_{abc}^{(0)}$ coefficients uniquely. 

Under conditions \eqref{sec32_3}, from eq. \eqref{sec32_2} we can read the two trace currents 
\begin{equation}
\Theta^{(\pm)}=-\frac{1}{2} b_{ab} V_b^{(\pm)}\partial \phi^{(\pm)}_a - \frac{1}{2}c_{ab} V^{(\pm)}_b \partial \phi^{(\pm)}_a - \frac{\gamma}{8} c_{ab} \partial V^{(\pm)}_{ab} - \frac{\gamma^2}{48} a_{abc} \partial V^{(\pm)}_{abc} 
\label{sec32_6}
\end{equation}
An analogous calculation leads to the $\tilde{J}^{(\pm)}$ and $\tilde{\Theta}^{(\pm)}$ currents, which are simply given by the previous expressions where we replace $\partial \to \bar{\partial}$. 

Before placing the defect at $x_1=0$, the present conservation law leads to spin-$(\pm 2)$ conserved charges, which can be written in the form \eqref{intro8.1} or, equivalently, in the energy-like and momentum-like forms \eqref{intro11}. However, as discussed in section \ref{classical}, the defect breaks this degeneracy and already at classical level only a linear combination of $Q_{+2}$ and $Q_{-2}$ survives, which is determined by the particular choice of the defect couplings.  In order to carry out the quantum analysis in a definite case, we choose defect \eqref{sec2_14bis} which, for equal root bases, ensures conservation of the $Q_{+2}$ charge at classical level, while discarding $Q_{-2}$.  

According to the general definitions of section \ref{procedure}, conservation of  $Q_{+2}$ at quantum level requires imposing the following condition  (see eq. \eqref{intro19})
\begin{equation}
\Bigl \langle J^{(+)} - J^{(-)}-\Theta^{(+)}+\Theta^{(-)} \Bigr \rangle\Bigr|_{x_1=0}  = \partial_0\hspace{-0.1cm}-\hspace{-0.1cm}{\rm derivative}
\label{sec322_1}
\end{equation}
We find convenient to rewrite this expression as a linear combination of energy- and momentum-like currents. In fact, as
explained in section \ref{procedure}, it then follows that non-trivial local contributions to the former expectation value  come from an odd power expansion in $S_{d}$, while the latter can be found expanding $S_{d}$ in even powers. Since the calculation is quite cumbersome, we report only the final result. Up to total $\partial_0$-derivatives arising from the classical part of the $c_{ab}$-terms in \eqref{sec32_1}, we obtain\footnote{The reader should be aware that in \eqref{sec322_8} the subscripts have different meaning according to which quantity they refer to: $R_a$ means the $a$-component of the vector field \eqref{sec321_5}, whereas $B_{ab \dots}$ and $V_{ab \dots}$ stay for derivatives of the $B,V$ functions respect to $\phi_a, \phi_b, \dots $.}
\begin{equation}\begin{split} \label{sec322_8}
& \Bigl \langle J^{(+)} - J^{(-)}-\Theta^{(+)}+\Theta^{(-)} \Bigr \rangle\Bigr|_{x_1=0} = \\
&\frac{i}{2\sqrt{2}}a_{abc} \,  (B^{+}_c+B^{-}_c)  \partial_0R_a\partial_0R_b
+  i \sqrt{2}b_{ab} \,   \partial_0 \bigl(B_a^{+}+B_a^{-}\bigr)  \partial_0R_b  +\frac{i\gamma}{\sqrt{2}}a_{abc} \,  \partial_0 \bigl( B_{bc}^{++}+B_{bc}^{--} \bigr)  \partial_0R_a \\
&-\frac{1}{2\sqrt{2}}a_{abc} \,  \bigl(B_b^{+}B_c^{+}-B_b^{-}B_c^{-}\bigr) \partial_0R_a
-\frac{1}{\sqrt{2}}b_{ab}\Bigl[\partial_0B_a^{+} B_b^{+}-\partial_0B_a^{-} B_b^{-} + \partial_0R_b\bigl(V_a^{(+)}-V_a^{(-)} \bigr)\Bigr]\\
&+\frac{\gamma}{8\sqrt{2}} c_{lb}\Bigl( V_{lba}^{(+)}-V_{lba}^{(-)} \Bigr) \partial_0R_a-\frac{\gamma}{\sqrt{2}}a_{abc} \Bigl( B_a^+ \partial_0B_{bc}^{++}-B_a^- \partial_0B_{bc}^{--} \Bigr) \\
&-\frac{i}{6\sqrt{2}}a_{abc}\Bigl(B_a^{+} B_b^{+}B_c^{+}+B_a^{-}B_b^{-}B_c^{-}\Bigr)\\
&-\frac{i}{\sqrt{2}}b_{ab}\Bigl(V_a^{(+)}B_b^{+}+V_a^{(-)}B_b^{-}\Bigr)+i\frac{\gamma}{8\sqrt{2}}c_{ab}\Bigl(V_{abc}^{(+)}B_c^{+}+V_{abc}^{(-)}B_c^{-}\Bigr) 
\end{split}\end{equation}
where, as done for the sinh-Gordon case, we have grouped the terms according to the number of $\partial_0$-derivatives. In principle a classical term proportional to three $\partial_0$-derivatives should be also present, but it vanishes once we require $E=F=1-D$.

Setting $\gamma=0$ we are left with an expression that reduces to a total $\partial_0$-derivative if $B$ is the classical potential \eqref{sec21_14} and the $D$-matrix hidden inside the $R_a$ fields is exactly the one given in \eqref{sec2_14bis} (see appendix \ref{App:spin3} for the detailed proof). 

For $\gamma \neq 0$ this is no longer true and we need to look for a possible quantum redefinition of the defect potential which makes this expression a total $\partial_0$-derivative. We then set 
\begin{equation}\begin{split}
& B = B^{(0)}+\gamma B^{(1)}+\gamma^2 B^{(2)} + \ldots  \\
& D=D^{(0)}+\gamma D^{(1)}+\gamma^2 D^{(2)} + \ldots , \qquad \qquad E=F = 1-D
\label{sec322_3}
\end{split}
\end{equation}
where $B^{(0)}$ the classical potential \eqref{sec21_14} and $D^{(0)}$ the classical matrix (\ref{sec2_14bis}). 

Inserting in \eqref{sec322_8} we end up with an expression  in which quantum corrections proportional to $\gamma$ not only appear explicitly, but are also hidden in the $B$-terms and in $R = D \phi^{(+)} + D^T \phi^{(-)}$, through definitions \eqref{sec322_3}.

First of all, we focus on the first line of \eqref{sec322_8} proportional to the second power of $\partial_0$-derivatives of the fields. Taking into account that at order $\gamma^0$ it vanishes if the defect potential is of the form in (\ref{sec21_14}) together with the fact that it is linear in $B$, the only possibility for it to vanish at all orders in $\gamma$ is by requiring the renormalized $B$ to be of the form
\begin{equation}
B=\sum_{i=0}^r\Bigl(l_i (\gamma) \, e^{\frac{1}{2}\alpha_i \cdot (D^{(0)} \phi^{(+)}+D^{(0)T} \phi^{(-)})}+s_i (\gamma) \, e^{\frac{1}{2}\alpha_i \cdot D^{(0)}(\phi^{(-)}-\phi^{(+)})} \Bigr)
\label{sec322_10}
\end{equation}
where $l_i (\gamma)= l_i^{(0)} + \gamma l_i^{(1)} + \dots$, $s_i (\gamma) = s_i^{(0)} + \gamma s_i^{(1)} + \dots$, and at classical level $l_i^{(0)} s_i^{(0)} = 1$. In other words, the interaction exponential terms must be the same as their classical counterparts, whereas quantum corrections can appear in the overall coefficients. 

We can then use the freedom to fix these coefficients in order to make the rest of the terms vanishing, up to total $\partial_0$-derivatives. We have performed the calculation for $A_{r}^{(1)}$ theories up to $r=5$, and unfortunately in all the cases we have found that there is no way to cancel completely expression \eqref{sec322_8}. The situation does not improve either by using conditions \eqref{sec21_13} that ensure momentum conservation, or smoothness conditions at the defect like \eqref{sec31_11bisbisbis}. 
Since we considered the most general situation (free renormalization of $D$, $c_{ab}$ and $B$) we conclude that the conservation law associated to the spin-$3$ current gets spoiled at quantum level. 

It is important to note that there is still one possibility that needs to be considered. This corresponds to setting $D=E=F=0$ in the  general expression of the defect potential \eqref{intro3} and redo the calculation all over again. We find that the energy-like $(Q_{+2} + Q_{-2})$ charge is conserved at quantum level if we choose a defect potential $B = B(\phi^{(+)}) + B(\phi^{(-)})$ with B given by  
\begin{equation}
B = \sqrt{1 + \frac{\gamma}{2}} \, \sum_{j=0}^r d_j e^{\frac12 \alpha \cdot \phi} , \qquad \quad d_j^2 = 4
\end{equation}
In this case the system reduces simply to the juxtaposition of two Toda theories with boundary \cite{a32,a34}.

\section{Conclusions}\label{conclusions}

We have investigated quantum integrability of $A_r^{(1)}$ Toda theories in the presence of a type-I defect. This has been done by studying directly the conservation laws of spin-$(s+1)$ currents in the bulk and at the defect and searching for possible quantum redefinitions of the currents and the defect potential that ensure anomalies cancellation. We have applied massless perturbation theory which, although perturbative in spirit, allows to obtain exact all-order results. 

For all the theories we have proved that the total energy is always conserved without need of quantum redefinitions. In particular, the exact expression of the energy emitted by the defect has been determined. In addition,   if we apply a suitable renormalization of the defect potential also the total momentum is conserved, with a non-trivial injection of momentum from the defect. 

Unfortunately, the situation gets worse when we study higher spin currents. In fact, the analysis of the spin-$4$ current in the sinh-Gordon model and the spin-$3$ current in $A_{r>1}^{(1)}$ theories has revealed that there is no finite renormalization of the defect potential which may cancel quantum anomalies. Therefore, these conservation laws get irreparably spoiled at quantum level. A couple of degenerate cases make an exception.  These correspond to the case in which the system reduces to a pair of decoupled bulk theories with boundary, and the sinh-Gordon case in which the left and right fields are constrained on the defect to differ by a constant.

Even though our negative results are not sufficient to conclude that integrability is lost at quantum level, it is quite hard to believe that we may find a better behavior in higher spin currents. In fact, higher is the spin of the current higher is the number of conditions that we should impose in order to cancel anomalies, thus rendering the conservation more and more doubtful. 

At quantum level the existence of higher spin conserved currents has dramatic effects on the structure of the S-matrix. In fact, as a consequence, it factorizes into the product of elastic two-particle S-matrices that can be determined exactly imposing analyticity, unitarity, crossing symmetry, Yang-Baxter identities and the bootstrap principle \cite{a3,a4,a16,a17,a18,a19,a20,aa20,a21,a22,a23,a24,aa24,a25,a26,Delius:1990ij,Delius:1991sv}. Here we are facing a situation where this approach is still valid in the bulk but likely breaks near the defect. Therefore, asymptotic states can be constructed in the left and right theories, which undergo an integrable scattering as long as they are far away from the defect. It would be interesting to investigate what happens when these particles hit the defect, for instance along the lines of \cite{b25,b26}.

We have worked with the assumption for the $\beta$ coupling to be  real. However, our results can be easily generalized to complex Toda theories. For instance, recovering the coupling in the exponential potentials by the replacement $\phi \to \beta \phi$ and then sending $\beta \to i \beta$ the sinh-Gordon model (\ref{sec31_1}, \ref{sec31_4}) maps to the sine-Gordon model with a line of defect. It is then easy to generalize our results to this case. We still find anomalies in the conservation laws that cannot be cancelled by quantum redefinitions of the defect potential, so leading to doubtful integrable properties of the quantum theory. 
On the other hand, in \cite{b19,b20} assuming the theory to be integrable an exact defect transmission matrix has been proposed, which reproduces correctly all the classical transmission properties that can be ascribed to the sine-Gordon theory. A possible interpretation of these two contradictory results is that the usual sine-Gordon lagrangian with type-I defect is not the correct lagrangian to implement quantum computations. In other words, the trasmission matrix found in \cite{b19,b20} describes presumably a theory that still leaks a correct lagrangian formulation at quantum level or a quantum theory which is non-lagrangian at all.
A perturbative computation of the soliton-defect scattering could shed some light on the role of the anomalies in the bootstrap approach. If everything is consistent we expect the appearance of $\beta^2$-corrections that invalidate the triangle equations. 

More recently type-II line defects, i.e. defects with additional degrees of freedom $\lambda$ propagating on them, have been investigated from a classical point of view \cite{b22, b23, b24, b27, b28}. It has been argued that a larger class of integrable theories seem to survive the presence of this kind of more general impurities. In the particular case in which the potential $B$ is independent of $\lambda$, these additional fields act as lagrangian multipliers forcing the condition $\partial_0 \phi^{(-)} = \partial_0 \phi^{(+)}$ at the defect. Interestingly, this is exactly condition \eqref{sec31_11bisbisbis} that ensures conservation of the spin-4 current in the sinh-Gordon model. Although this is quite promising, the whole calculation should be redone in the presence of these additional fields, as new $\lambda$-dependent derivative terms are expected to arise, which would change the expression of the anomalies. 

It would be interesting to generalize our analysis to nonsimply-laced Toda theories with a line of defect. Quantum integrability of these theories has been studied so far in the bulk \cite{a12} and in the presence of a boundary \cite{a33,a34}. In the more general framework of type-II impurities there are strong signs that these theories are still classically integrable \cite{Robertson:2013cna,b27}, despite a quantum analysis of the problem has not been performed yet.

Another interesting application would be the study of supersymmetric Toda models based on superalgebras. For theories defined on the whole line the quantum conservation laws have been studied in \cite{a14} where a suitable renormalization of the currents preserving conservation laws has been found. Classically it is possible to introduce boundary \cite{Aguirre:2013pja} and defects \cite{c1,c2,c3,c4,c5} but a systematic classification of possibly susy preserving boundaries and defects at quantum level is still lacking. 

Finally, our approach could be used for studying Janus configurations in two dimensions, that is defect theories characterized by different values of the coupling constant on the two sides of the defect \cite{b30,b31}.

\vskip 40pt
\noindent
{\bf Acknowledgments}

\noindent
We acknowledge stimulating discussions with Peter Bowcock. This work has been supported in part by MIUR -- Italian Ministero dell'Istruzione, Universit\`a e Ricerca, and by INFN -- Istituto Nazionale di Fisica Nucleare within the ``Gauge Theories, Strings, Supergravity'' (GSS) research project. This project has received funding from the  European Union's  Horizon  2020  research  and  innovation programme under the Marie  Sk\l odowska-Curie  grant  agreement  No.  764850 \textit{``SAGEX''}.

\newpage

\appendix 
\section{Computation of momentum flow on the defect} \label{App:momentumflow}

In this appendix we give an explicit example of calculation in which local contributions to the current flow at the defect come from higher-order powers in the $S_d$ expansion. The simplest case in which this happens is for the momentum flow, as described in section \ref{procedure}. Referring to the r.h.s. of eq. \eqref{sec21_5bis} we focus on the first expectation value in the second line. Expanding around the trivial vacuum and using the explicit expression \eqref{intro3} for the defect potential we can write
\begin{equation}\begin{split}
&\Bigl \langle \partial \phi^{(+)}_a \partial \phi^{(+)}_a+\bar{\partial} \phi^{(+)}_a \bar{\partial} \phi_a^{(+)} \Bigr \rangle = \Bigl \langle \partial \phi^{(+)}_a \partial \phi^{(+)}_a+\bar{\partial} \phi^{(+)}_a \bar{\partial} \phi_a^{(+)} \Bigr \rangle_{\hspace{-0.1cm}0}\\
&+\frac{1}{2\beta^4}\int \int d\omega_0 d\tilde{\omega}_0\biggl \langle \Bigl( \partial \phi^{(+)}_a \partial \phi^{(+)}_a+\bar{\partial} \phi^{(+)}_a \bar{\partial} \phi_a^{(+)} \Bigr)(x)\\
&\Bigl( \frac{i}{2} \phi_b^{(+)} F_{bc} \partial_0 \phi_c^{(+)} +i \phi_b^{(-)} D_{bc} \partial_0 \phi_c^{(+)} + B  \Bigr) (\omega_0) \Bigl( \frac{i}{2} \phi_d^{(+)} F_{de} \partial_0 \phi_e^{(+)} +i \phi_d^{(-)} D_{de} \partial_0 \phi_e^{(+)} + B  \Bigr) (\tilde{\omega}_0)\biggr \rangle_{\hspace{-0.1cm}0}
\label{Amom1}
\end{split}\end{equation}
where we have omitted the first order term $S_d$ since it does not contribute with local terms, as already discussed. 
Performing all possible Wick contractions and taking the limit at the defect, we obtain
\begin{equation}\begin{split}
&\Bigl \langle \partial \phi^{(+)}_a \partial \phi^{(+)}_a+\bar{\partial} \phi^{(+)}_a \bar{\partial} \phi_a^{(+)} \Bigr \rangle \Bigl|_{x_1 = 0} = \partial_0 \phi_a^{(+)} \partial_0 \phi_a^{(+)} \Bigl|_{x_1 = 0}  \\
&+\lim_{x_1\to 0^+}\frac{1}{4\pi^2} \int d\omega_0 d\tilde{\omega}_0 \biggl( \frac{1}{(x_0-\omega_0 - ix_1)(x_0-\tilde{\omega}_0 - ix_1)}+\frac{1}{(x_0-\omega_0 + ix_1)(x_0-\tilde{\omega}_0 + ix_1)}  \biggr)  \times\\
&  \biggl( -2 F_{ac} \partial_0\phi_c^{(+)}(\omega_0) F_{ae}\partial_0\phi_e^{(+)}(\tilde{\omega}_0) +4 F_{ac} \partial_0\phi_c^{(+)}(\omega_0) \partial_0\phi_d^{(-)}(\tilde{\omega}_0) D_{da}  +4i F_{ac} \partial_0\phi_c^{(+)}(\omega_0) B^+_a(\tilde{\omega}_0) \\
& \qquad -2 \partial_0\phi_b^{(-)}(\omega_0) D_{ba}\partial_0\phi_d^{(-)}(\tilde{\omega}_0) D_{da}
 -4i \partial_0\phi_b^{(-)}(\omega_0) D_{ba} B_a^+(\tilde{\omega}_0)+2B_a^+(\omega_0)B_a^+(\tilde{\omega}_0)   \biggr)
\label{Amom2}
\end{split}\end{equation}
From this expression local contributions can arise only if in the $x_1 \to 0^+$ limit we produce two delta functions that remove the $\omega_0, \tilde{\omega}_0$ integrations. Using the general identity
\begin{equation}
\lim_{x_1 \to 0^+} \frac{1}{x_0-\omega_0\mp ix_1} = Pv \frac{1}{x_0-\omega_0} \pm i \pi \delta(x_0-\omega_0)
\label{Amom3}
\end{equation}
in \eqref{Amom2} we can select
\begin{equation} \begin{split}
& \lim_{x_1\to 0^+} \biggl( \frac{1}{(x_0-\omega_0 - ix_1)(x_0-\tilde{\omega}_0 - ix_1)}+\frac{1}{(x_0-\omega_0 + ix_1)(x_0-\tilde{\omega}_0 + ix_1)}  \biggr) \\
& \longrightarrow  - 2 \pi^2 \delta (x_0-\omega_0) \delta (x_0-\tilde{\omega}_0)  
\end{split}
\end{equation} 
Then performing the integrals, up to non-local terms we obtain
\begin{equation}\begin{split}
&\Bigl \langle \partial \phi^{(+)}_a \partial \phi^{(+)}_a+\bar{\partial} \phi^{(+)}_a \bar{\partial} \phi_a^{(+)} \Bigr \rangle \Bigl|_{x_1=0}=  \\
&\partial_0\phi^{(+)} \Bigl(1 +F^T F\Bigr) \partial_0\phi^{(+)}+ \partial_0\phi^{(-)} D D^T  \partial_0\phi^{(-)} - 2  \partial_0\phi^{(-)} D F  \partial_0\phi^{(+)}\\
&+2i \partial_0\phi^{(-)}D B^+ -2i \partial_0 \phi^{(+)} F^T B^+ - (B^+)^2
\end{split}\end{equation}
We note that this is an entirely classical contribution, which could have been alternatively obtained by simply substituting the interacting boundary equations (\ref{intro13}) on the left hand side of expression (\ref{Amom1}). This is a crossed-check of our procedure.

 An analogous calculation can be done for the $\phi^{(-)}$ term in \eqref{sec21_5bis}. Summing the two contributions we obtain the momentum flow in (\ref{sec21_11}). \\

\section{Classical spin-2 charges} \label{App:spin3}

In this appendix we study in details the classical conservation law for spin-($\pm2$) charges. As recalled in the main text, this is a particularly instructive case since already at classical level the defect breaks degeneracy and only a linear combination of $Q_{+2}$ and $Q_{-2}$ turns out to be conserved. 
This has been previously proved in \cite{b18} using the classical equations of motion. Here we provide an alternative derivation using the quantum calculations of section \ref{sect:spin3} where we set $\gamma=0$. 

We consider the classical conserved spin-3 currents in the bulk
\begin{equation}\begin{split} \label{BAppendix1}
& J^{(\pm)}=\frac{1}{3}a_{abc}\partial \phi_a^{(\pm)} \partial \phi_b^{(\pm)} \partial \phi_c^{(\pm)} + b_{ab} \partial^2\phi^{(\pm)}_a \partial \phi^{(\pm)} _b \\
& \Theta^{(\pm)}=-\frac{1}{2} b_{ab} V_b^{(\pm)}\partial \phi^{(\pm)}_a  
\end{split}\end{equation}
with the $a_{abc}$ and $b_{ab}$ coefficients satisfying constraint \eqref{sec32_4}. Without loosing generality we take the solution \cite{b18}
\begin{equation}
b_{ab}=\delta_{ab}-D_{ab}
\label{BAppendix5}
\end{equation}
$\tilde{J}^{(\pm)}$ and $\tilde{\Theta}^{(\pm)}$ currents are simply obtained by substituting $\partial \to \bar{\partial}$ in the previous expressions. Without the defect, these currents lead to two conserved charges, $Q_{+2}$ and $Q_{-2}$, of opposite spin \cite{a1}.

We then study the classical conservation laws \eqref{intro19} at the defect. We apply the general procedure described in section \ref{sect2} by taking into account that classical contributions will come only from terms where the number of field contractions equals the order in the expansion of $e^{\frac{1}{\beta^2}(V^{(+)} + V^{(-)} - V^{(d)})}$. The result can be expressed as a linear combination of terms grouped according to the number of $\partial_0$-derivatives of the fields.  

We begin discussing terms proportional to cubic powers in the field derivatives, which for the $Q_{+2}$ charge read
\begin{equation}\begin{split} \label{BAppendix5.1}
&\Bigl ( J^{(+)} - J^{(-)}-\Theta^{(+)}+\Theta^{(-)} \Bigr )\Bigr|_{x_1=0}^{(3)} =\\
&+\frac{1}{6\sqrt{2}} a_{abc} \Bigl( D^{T}_{ad} D^{T}_{be} D^{T}_{cf} - (1+E)_{ad} (1+E)_{be} (1+E)_{cf} \Bigr)\partial_0\phi_d^{(-)}\partial_0\phi_e^{(-)}\partial_0\phi_f^{(-)}\\
&+\frac{1}{2\sqrt{2}} a_{abc} \Bigl( D^{T}_{ad} (1-F)_{be} (1-F)_{cf} - (1+E)_{ad} D_{be} D_{cf} \Bigr)\partial_0\phi_d^{(-)}\partial_0\phi_e^{(+)}\partial_0\phi_f^{(+)}\\
&+\frac{1}{2\sqrt{2}} a_{abc} \Bigl( D^{T}_{ad} D^T_{be} (1-F)_{cf} - (1+E)_{ad} (1+E)_{be} D_{cf} \Bigr)\partial_0\phi_d^{(-)}\partial_0\phi_e^{(-)}\partial_0\phi_f^{(+)}\\
&+\frac{1}{6\sqrt{2}} a_{abc} \Bigl( (1-F)_{ad} (1-F)_{be} (1-F)_{cf} - D_{ad} D_{be} D_{cf} \Bigr)\partial_0\phi_d^{(+)}\partial_0\phi_e^{(+)}\partial_0\phi_f^{(+)}\\
&+\frac{1}{\sqrt{2}} b_{ab} \Bigl( D^{T}_{ac} D^{T}_{bd} - (1+E)_{ac} (1+E)_{bd} \Bigr)\partial^2_0\phi_c^{(-)}\partial_0\phi_d^{(-)}\\
&+\frac{1}{\sqrt{2}} b_{ab} \Bigl( D^{T}_{ac} (1-F)_{bd} - (1+E)_{ac} D_{bd} \Bigr)\partial^2_0\phi_c^{(-)}\partial_0\phi_d^{(+)}\\
&+\frac{1}{\sqrt{2}} b_{ab} \Bigl( (1-F)_{ac} D^{T}_{bd} - D_{ac} (1+E)_{bd} \Bigr)\partial^2_0\phi_c^{(+)}\partial_0\phi_d^{(-)}\\
&+\frac{1}{\sqrt{2}} b_{ab} \Bigl( (1-F)_{ac} (1-F)_{bd} - D_{ac} D_{bd} \Bigr)\partial^2_0\phi_c^{(+)}\partial_0\phi_d^{(+)}\\
\end{split}\end{equation}
whereas for $Q_{-2}$
\begin{equation}\begin{split} \label{BAppendix5.2}
&\Bigl ( \tilde{J}^{(-)} - \tilde{J}^{(+)} + \tilde{\Theta}^{(+)} - \tilde{\Theta}^{(-)} \Bigr )\Bigr|_{x_1=0}^{(3)} =\\
&+\frac{1}{6\sqrt{2}} a_{abc} \Bigl( D^{T}_{ad} D^{T}_{be} D^{T}_{cf} + (1-E)_{ad} (1-E)_{be} (1-E)_{cf} \Bigr)\partial_0\phi_d^{(-)}\partial_0\phi_e^{(-)}\partial_0\phi_f^{(-)}\\
&+\frac{1}{2\sqrt{2}} a_{abc} \Bigl( D^{T}_{ad} (1+F)_{be} (1+F)_{cf} + (1-E)_{ad} D_{be} D_{cf} \Bigr)\partial_0\phi_d^{(-)}\partial_0\phi_e^{(+)}\partial_0\phi_f^{(+)}\\
&-\frac{1}{2\sqrt{2}} a_{abc} \Bigl( D^{T}_{ad} D^T_{be} (1+F)_{cf} + (1-E)_{ad} (1-E)_{be} D_{cf} \Bigr)\partial_0\phi_d^{(-)}\partial_0\phi_e^{(-)}\partial_0\phi_f^{(+)}\\
&-\frac{1}{6\sqrt{2}} a_{abc} \Bigl( (1+F)_{ad} (1+F)_{be} (1+F)_{cf} + D_{ad} D_{be} D_{cf} \Bigr)\partial_0\phi_d^{(+)}\partial_0\phi_e^{(+)}\partial_0\phi_f^{(+)}\\
&+\frac{1}{\sqrt{2}} b_{ab} \Bigl( (1-E)_{ac} (1-E)_{bd}-D^{T}_{ac} D^{T}_{bd} \Bigr)\partial^2_0\phi_c^{(-)}\partial_0\phi_d^{(-)}\\
&+\frac{1}{\sqrt{2}} b_{ab} \Bigl(D^T_{ac} (1+F)_{bd}-(1-E)_{ac}  D_{bd} \Bigr)\partial^2_0\phi_c^{(-)}\partial_0\phi_d^{(+)}\\
&+\frac{1}{\sqrt{2}} b_{ab} \Bigl((1+F)_{ac} D^T_{bd} - D_{ac} (1-E)_{bd} \Bigr)\partial^2_0\phi_c^{(+)}\partial_0\phi_d^{(-)}\\
&+\frac{1}{\sqrt{2}} b_{ab} \Bigl(D_{ac} D_{bd} - (1+F)_{ac} (1+F)_{bd} \Bigr)\partial^2_0\phi_c^{(+)}\partial_0\phi_d^{(+)}\\
\end{split}\end{equation}
Conservation requires the two expressions to vanish, up to a total $\partial_0$-derivative. However, these two conditions are incompatible, so they discriminate between the two charges. In fact, if we choose $E=F=1-D$ then expression \eqref{BAppendix5.1} vanishes while \eqref{BAppendix5.2} does not, and $Q_{-2}$ is lost. Alternatively, if we choose $E=F=-1-D$ the opposite situation occurs and we loose $Q_{+2}$. These two options are both solutions of (\ref{sec21_11.1}) and therefore they are connected by a transformation of the form in (\ref{sec21_11.2}). Interestingly, each of them is an eligible solution for a defect potential that preserves energy and momentum.

In order to be consistent with what we have done in section \ref{classical} we choose solution $E=F=1-D$ that affects the $Q_{-2}$ conservation. The rest of this section is devoted to prove that $Q_{+2}$ is indeed conserved. 

Apart from cubic terms in the field derivatives that we have already discussed, the rest of classical contributions can be read directly from result  \eqref{sec322_8} by setting $\gamma=0$.  Terms proportional to two $\partial_0$-derivatives are
\begin{equation}\begin{split}
&\Bigl( J^{(+)} - J^{(-)}-\Theta^{(+)}+\Theta^{(-)} \Bigr)\Bigl|_{x_1=0}^{(2)}\\
&\qquad = \frac{i}{2\sqrt{2}}a_{abc} (B^{+}_c+B^{-}_c)\partial_0R_a\partial_0R_b+i\sqrt{2} b_{ab} \partial_0 \bigl(B_a^{+}+B_a^{-}\bigr)  \partial_0R_b\\
&\qquad = \frac{i}{2\sqrt{2}}\partial_0R_a \partial_0R_b \sum_{i=1}^r \frac{1}{\sigma} e^{\frac{1}{2}\alpha_i \cdot R} \Bigl[ a_{abc} (\alpha_i)_c +2 b_{cb} (\alpha_i)_c (\alpha_i)_a\Bigr]
\label{BAppendix7}
\end{split}\end{equation}
where the last equality has been obtained using the structure for the defect potential in  (\ref{sec21_14}) and (\ref{eq:sec21_15a}). Now, thanks to relation (\ref{sec32_4}) among the current coefficients it is easy to see that this expression vanishes identically.

Lower order terms in the $\partial_0$-derivatives are more difficult to deal with and require more algebra. Exploiting the symmetry of $a_{abc}$, the first order contributions acquire the following form
\begin{equation}\begin{split}
&\Bigl( J^{(+)} - J^{(-)}-\Theta^{(+)}+\Theta^{(-)} \Bigr)\Bigl|_{x_1=0}^{(1)}=-\frac{1}{2\sqrt{2}}a_{abc}\bigl(B_b^+ - B_b^-\bigr) \bigl(B_c^++B_c^-\bigr)\partial_0R_a\\
&-\frac{1}{\sqrt{2}}b_{ab}\Bigl[\partial_0B_a^+ B_b^+ -\partial_0B_a^- B_b^- \Bigr] -\frac{1}{\sqrt{2}}b_{ab} \partial_0R_b\bigl(V_a^{(+)}-V_a^{(-)} \bigr)
\label{BAppendix8}
\end{split}\end{equation}
This expression can be evaluated by inserting the explicit expressions for the $V^{(\pm)}$ and $B$ potentials in \eqref{intro1} and \eqref{sec21_14} respectively, plus solution \eqref{BAppendix5} for the $b_{ab}$ coefficients. It can be further elaborated by exploiting identities (\ref{eq:sec21_15a}, \ref{eq:sec21_15b}) which also imply $DbD^T=D^TbD$ and
\begin{equation}
\label{BAppendix10}
\alpha_i (1-D)D^T\alpha_j=\alpha_jD(1-D^T)\alpha_i=-\alpha_jD(1-D)\alpha_i 
\end{equation}
Summing all the contributions, there are some cancellations and we are left with 
\begin{equation}\begin{split}
&\Bigl( J^{(+)} - J^{(-)}-\Theta^{(+)}+\Theta^{(-)} \Bigr)\Bigl|_{x_1=0}^{(1)}=\frac{1}{2\sqrt{2}}\partial_0\biggl[\sum_{i,j=0}^r \alpha_jD(1-D)\alpha_i e^{\frac{1}{2}\alpha_i \cdot R} e^{\frac{1}{2}\alpha_j \cdot S}\biggr]\\
&+\frac{1}{2\sqrt{2}\sigma^2}\sum_{i,j=0}^r \Bigl[ \bigl(\alpha_j (1-D)^2\alpha_i\bigr)  \alpha_i \cdot \partial_0R - \bigl(\alpha_j(1-D)\alpha_i\bigr)\bigl( \alpha_i (1-D) \partial_0R\bigr) \Bigr] e^{\frac{1}{2} (\alpha_i+\alpha_j) \cdot R} 
\label{BAppendix14}
\end{split}\end{equation}
The last task is to check whether the second line is also a total $\partial_0$-derivative. This will be the case if every single exponential $e^{\frac{1}{2}(\alpha_i+\alpha_j)\cdot R}$  times its relative multiplicative factor reduces to a total derivative. 

For the first term we can exploit the identity $\alpha_j(1-D)^2\alpha_i=\alpha_i(1-D)^2\alpha_j$ which follows from the antisymmetry of $(1-D)$ matrix and easily write
\begin{equation}
\sum_{i,j=0}^r  \bigl(\alpha_j (1-D)^2\alpha_i\bigr)  \alpha_i \cdot \partial_0R  \, e^{\frac{1}{2} (\alpha_i+\alpha_j) \cdot R}= \partial_0\Bigl[ \sum_{i,j=0}^r(\alpha_j(1-D)^2\alpha_i) e^{\frac{1}{2} (\alpha_i+\alpha_j) \cdot R} \Bigr]
\label{BAppendix15}
\end{equation}
For the second term we first exploit the following identity 
\begin{equation}
\alpha_j (1-D) \alpha_i=\delta_{j,i+1}-\delta_{j,i-1}
\end{equation}
which follows from \eqref{intro2} and \eqref{eq:sec21_15c}. This, together with the ciclicity property of the Dynkin diagram, allows to write 
\begin{equation}
-\sum_{i,j} \bigl(\alpha_j(1-D)\alpha_i\bigr)\bigl( \alpha_i (1-D)  \partial_0R\bigr)  \, e^{\frac{1}{2} (\alpha_i+\alpha_j) \cdot R}= \sum_{j} (\alpha_{j+1}-\alpha_j)(1-D)  \cdot \partial_0 R \, e^{\frac{1}{2}(\alpha_{j+1} + \alpha_j)\cdot R}
\label{BAppendix17bis}
\end{equation}
At this point properties (\ref{eq:sec21_15a}, \ref{eq:sec21_15b}) can be exploited to write
\begin{equation}\begin{split}
(\alpha_{j+1}-\alpha_j) (1-D) &=\alpha_{j+1}-\alpha_j-\alpha_{j+1}D+\alpha_jD\\
&=\alpha_{j+1}-\alpha_j+\alpha_j(D^T+D)=\alpha_j+\alpha_{j+1} 
\end{split}\end{equation}
so that each term in \eqref{BAppendix17bis} reduces to $2 \partial_0 e^{\frac{1}{2} (\alpha_{j+1}+\alpha_j) \cdot R} $.

In conclusion, the terms proportional to a single field derivative are a total derivative
\begin{equation}\begin{split}
&\Bigl( J^{(+)} - J^{(-)}-\Theta^{(+)}+\Theta^{(-)} \Bigr)\Bigl|_{x_1=0}^{(1)}=\\
& \frac{1}{2\sqrt{2}}\partial_0\biggl[\sum_{i,j=0}^r \alpha_jD(1-D)\alpha_i \; e^{\frac{1}{2}\alpha_i \cdot R} e^{\frac{1}{2}\alpha_j \cdot S} \\
&\qquad  \qquad +\frac{1}{\sigma^2}\sum_{i,j=0}^r \alpha_j(1-D)^2\alpha_i \; e^{\frac{1}{2} (\alpha_i+\alpha_j) \cdot R} 
 +\frac{2}{\sigma^2}\sum^r_{j=0} e^{\frac{1}{2}(\alpha_{j+1} + \alpha_j)\cdot R} \biggr]
\label{BAppendix17}
\end{split}\end{equation}

Finally we need to consider contributions with no powers of field derivatives. They are given by
\begin{equation}\begin{split}
&\Bigl( J^{(+)} - J^{(-)}-\Theta^{(+)}+\Theta^{(-)} \Bigr)\Bigl|_{x_1=0}^{(0)}\\
&= -\frac{i}{6\sqrt{2}}a_{abc}\bigl(B_a^+ B_b^+ B_c^+ + B_a^- B_b^- B_c^- \bigr)-\frac{i}{\sqrt{2}}b_{ab}\bigl(V_a^{(+)}B_b^+ + V_a^{(-)}B_b^-\bigr) \\
& \equiv \frac{1}{\sigma^3} X_1 + \frac{1}{\sigma}  X_2 + \sigma X_3 + \sigma^3 X_4
\label{BAppendix18}
\end{split}\end{equation}
where, recalling that the defect potential is a function of the free parameter $\sigma$, in the last line we have organized the result in powers of $\sigma$. It is then sufficient to prove that each $X_i$ is equal to zero, up to total derivatives. 

We start considering the term proportional to $1/\sigma^{3}$, which is explicitly given by 
\begin{equation}
X_1= -\frac{i}{48\sqrt{2}}a_{abc} \sum_{i,j,k=0}^r e^{\frac{1}{2}(\alpha_i+\alpha_j+\alpha_k) \cdot R} \Bigl( (\alpha_i D)_a(\alpha_j D)_b(\alpha_k D)_c+(\alpha_i D^T)_a(\alpha_j D^T)_b(\alpha_k D^T)_c\Bigr)
\end{equation}
This expression can be manipulated by using relations (\ref{eq:sec21_15a}) and (\ref{sec32_4}). It is straightforward to obtain
\begin{equation}\begin{split}
X_1 &=-\frac{i}{3 \sqrt{2}} \sum_{i,j,k=0}^r e^{\frac{1}{2}(\alpha_i+\alpha_j+\alpha_k) \cdot R} \Bigl[8 \alpha_i \cdot \alpha_k (\alpha_j (1-D) \alpha_k) 
\\
& \qquad \qquad \qquad \quad -3(\alpha_iD\alpha_k)\bigl(\alpha_jD^T(1-D)\alpha_k+\alpha_jD(-1+D^T)\alpha_i\bigr) \Bigr]
\label{BAppendix19}
\end{split}\end{equation}
Now, using the scalar product conventions (\ref{intro2}) and property (\ref{eq:sec21_15c}), thanks to the Kronecker deltas which arise we can reduce it to 
\begin{equation}\begin{split}
X_1 &= 4 \sqrt{2} i  \sum_{k=0}^r  \left[ e^{\frac{1}{2}(2\alpha_k+\alpha_{k-1}) \cdot R} - e^{\frac{1}{2}(2\alpha_k+\alpha_{k+1}) \cdot R} \right] 
+\sqrt{2} i \sum_{j,k=0}^r e^{\frac{1}{2}(2\alpha_k+\alpha_{j}) \cdot R}\bigl(\alpha_jD^T\alpha_k-\alpha_jD\alpha_k\bigr)\\
& \quad -\sqrt{2} i  \sum_{j,k=0}^r e^{\frac{1}{2}(\alpha_k+\alpha_{k+1}+\alpha_{j}) \cdot R}\bigl(\alpha_jD^T(1-D)\alpha_k+\alpha_jD(-1+D^T)\alpha_{k+1}\bigr)
\label{BAppendix20}
\end{split}\end{equation}
Finally, identities (\ref{eq:sec21_15a}-\ref{eq:sec21_15c}) allow to write
\begin{equation}\begin{split}
& \alpha_jD^T\alpha_k-\alpha_jD\alpha_k = 2  \bigl( \alpha_j \cdot \alpha_k -  \alpha_j D \alpha_k \bigr) = 2 \bigl(\delta_{j,k+1} - \delta_{j,k-1} \bigr)\\
&\alpha_jD^T(1-D)\alpha_k+\alpha_jD(-1+D^T)\alpha_{k+1}=\alpha_jD^T\alpha_k-\alpha_jD\alpha_{k+1}-2\alpha_jD\alpha_k \\
& \hspace{6.5cm} = 2(-\delta_{jk}-\delta_{j,k-1}+\delta_{j,k+1}+\delta_{j,k+2})
\end{split}\end{equation}
and inserting these results in \eqref{BAppendix20} it is easy to see that $X_1=0$, due to mutual cancellations. 

A similar procedure and the same kind of identities easily lead to $X_2=X_3=0$. Finally, the vanishing of $X_4$ is straightforward once we plug in $\bigl(B_a^+ B_b^+ B_c^+ + B_a^- B_b^- B_c^- \bigr)$ in eq. \eqref{BAppendix18} the explicit expression for the $B^{\pm}_a$ derivatives. 

This concludes the proof of the conservation of the $Q_{+2}$ charge at classical level. It is remarkable to note that the proof is highly non-trivial and works only thanks to a precise fine-tuning of the defect potential.

\end{document}